\documentclass[12pt,preprint]{aastex}

\slugcomment{Accepted for publication on the Astrophysical Journal}
\shorttitle{Solar Modulation using a Monte Carlo Dynamical Approach}
\shortauthors{P. Bobik et al.}

\begin{document}

%% LaTeX will automatically break titles if they run longer than
%% one line. However, you may use \\ to force a line break if
%% you desire.

\title{Systematic Investigation of Solar Modulation of Galactic Protons
for Solar Cycle 23 %and Helium-nuclei
using a Monte Carlo Approach with Particle Drift Effects and %Solar Wind
Latitudinal Dependence}

%% Use \author, \affil, and the \and command to format
%% author and affiliation information.
%% Note that \email has replaced the old \authoremail command
%% from AASTeX v4.0. You can use \email to mark an email address
%% anywhere in the paper, not just in the front matter.
%% As in the title, use \\ to force line breaks.

\author{P. Bobik\altaffilmark{1}, G. Boella\altaffilmark{2,3},
M.J. Boschini\altaffilmark{2,4}, C. Consolandi\altaffilmark{2}, S. Della Torre\altaffilmark{2,5}, M.
Gervasi\altaffilmark{2,3},\\ D. Grandi\altaffilmark{2}, K. Kudela\altaffilmark{1}, %E. Memola\altaffilmark{2},
S. Pensotti\altaffilmark{2,3}, P.G. Rancoita\altaffilmark{2} and M. Tacconi\altaffilmark{2}
}
\affil{Istituto Nazionale di Fisica Nucleare, INFN, Milano-Bicocca, Milano (Italy), I20126}

\email{piergiorgio.rancoita@mib.infn.it}

%% Notice that each of these authors has alternate affiliations, which
%% are identified by the \altaffilmark after each name.  Specify alternate
%% affiliation information with \altaffiltext, with one command per each
%% affiliation.

\altaffiltext{1}{Institute of Experimental Physics, Kosice (Slovak
Republic).}

\altaffiltext{2}{INFN, Milano-Bicocca, Milano (Italy).}

\altaffiltext{3}{also Physics Department, University of Milano-Bicocca, Milano (Italy).}

\altaffiltext{4}{also CILEA, Segrate (Milano, Italy).}

\altaffiltext{5}{also University of Insubria, Como (Italy).}

%% Mark off your abstract in the ``abstract'' environment. In the manuscript
%% style, abstract will output a Received/Accepted line after the
%% title and affiliation information. No date will appear since the author
%% does not have this information. The dates will be filled in by the
%% editorial office after submission.

\begin{abstract}
A propagation model of galactic cosmic protons through the Heliosphere was implemented using a 2-D Monte Carlo approach to
determine the differential intensities of protons during the solar cycle 23.~The model includes the effects due to the variation of
solar activity during the propagation of cosmic rays from the boundary of the
heliopause down to Earth's position.~Drift effects are also accounted for.~The simulated spectra were found in agreement with those obtained with experimental observations carried out by BESS, AMS and PAMELA collaborations.~In addition, the modulated spectrum determined with the present code for the year 1995 exhibits the latitudinal gradient and equatorial southward offset minimum found by Ulysses fast scan in 1995.

\end{abstract}

%% Keywords should appear after the \end{abstract} command. The uncommented
%% example has been keyed in ApJ style. See the instructions to authors
%% for the journal to which you are submitting your paper to determine
%% what keyword punctuation is appropriate.


%% From the front matter, we move on to the body of the paper.
%% In the first two sections, notice the use of the natbib \cite
%% and \citet commands to identify citations.  The citations are
%% tied to the reference list via symbolic KEYs. The KEY corresponds
%% to the KEY in the \bibitem in the reference list below. We have
%% chosen the first three characters of the first author's name plus
%% the last two numeral of the year of publication as our KEY for
%% each reference.

%% Authors who wish to have the most important objects in their paper
%% linked in the electronic edition to a data center may do so by tagging
%% their objects with \objectname{} or \object{}.  Each macro takes the
%% object name as its required argument. The optional, square-bracket
%% argument should be used in cases where the data center identification
%% differs from what is to be printed in the paper.  The text appearing
%% in curly braces is what will appear in print in the published paper.
%% If the object name is recognized by the data centers, it will be linked
%% in the electronic edition to the object data available at the data centers
%%
%% Note that for sources with brackets in their names, e.g. [WEG2004] 14h-090,
%% the brackets must be escaped with backslashes when used in the first
%% square-bracket argument, for instance, \object[\[WEG2004\] 14h-090]{90}).
%%  Otherwise, LaTeX will issue an error.
%
\section{Introduction}\label{Introduction}
\label{Intr}
During the last two decades - using balloon flights and space-borne missions -, the fluxes of
Galactic Cosmic Rays (GCR) and their energy distributions were observed in different phases of solar activity.~These data allow one to attempt a better understanding of processes related to the transport of GCRs through the Heliosphere.~Furthermore, the study of propagation properties - i.e., the effect of solar modulation on the fluxes - of GCRs may, in turn,
provide a tool to determine demodulated Local Interstellar
Spectra (LIS) of GCR components, for instance,
protons, light-nuclei, electrons,
positrons, anti-protons, etc., thus, a further understand of
processes of generation, acceleration and diffusion within the
Milky Way \citep[e.g., see][]{Boella,Galprop,DRAGON,Usine}.~In addition, an accurate determination of demodulated spectra may allow one
to untangle
features due to new physics - i.e.,~dark matter \citep[e.g., see][and references therein]{Bottino,Cirelli_2010,Ibarra_2010,Salati,Weniger} -  or
astrophysical sources \citep[e.g., see][and references therein]{ATIC, Fermi_elettroni,Pamela_Nature,Cern_Villa_O_2010,M_S_Villa_O_2010}.
\par
Recently, spectra of GCRs were obtained using dedicated spectrometers on space born-missions
\citep[e.g., see][]{AMS_protons,AMS_leptons,AMS_cosmic,AMS_helium,AMS_tot,AMS_positron,Pamela_Nature,Pamela_PRL,Pamela_AstroPH}
and balloon flights \citep[e.g., see][]{Caprice94,Imax92,BESS_PLB2,BESS_Astropart,BESS_PLB,BESS_ICRC}.~These spectra were measured i) with an accuracy
down to about or less than 30\% and ii) covering a time duration longer than a
solar cycle,~i.e., these spectra were measured under
solar conditions largely different.~These data can be hopefully
exploited to determine a general treatment of solar modulation in the inner heliosphere
to be used for different phases of solar activity and a better understanding of space radiation environment close to Earth
\citep[e.g., see][and references therein]{rop_si}.~In the near future even more accurate and systematic data will be available from AMS-02.~This spectrometer is
operational onboard of the International Space Station from May 2011 and is expected to collect data for more than a solar
cycle~\citep{Battiston2010,Davide2010}.~These observations will allow one to obtain accurate spectra with different solar activity conditions from some hundreds MeV up to very high energy (a few TeV's); in addition, using the same experimental apparatus, systematic errors on measured fluxes are expected to be minimized.~Furthermore,
observations made by the Ulysses spacecraft~\citep{simpson1992} in the inner
heliosphere could determine a latitudinal dependence of GCR (mostly protons) intensity with an equatorial southward offset minimum and a North polar excess~\citep[e.g., see][]{simpson1996b}.~Finally, it has be remarked that modulation phenomena were observed at low energies (i.e., lower than 500\,MeV/nucleon) in the outer heliosphere~\citep[e.g., see][]{webber2008} and are currently investigated, for instance, by~\citet{2003JGRA_108_8039L,2004JGRA_10901103L,reentrant2008,Potgieter_2008} (see also references therein).
\par
In the present model, a two dimensional (2-D) -~i.e., depending on the helio-colatitude and radial
distance from the Sun~\citep{Bobik03,Davide2008,Davide2010} -
Monte Carlo approach is adopted to solve the transport equation of propagation of GCRs down to the inner heliosphere, without addressing CR modulation observed in the outer.~The model exhibits
a slow time dependence because of the (almost) monthly averages of
solar activity parameters adopted for the i) solar wind speed ($V_{\rm sw}$), ii) tilt angle
($\alpha_{\rm t}$) of the neutral sheet and iii) diffusion parameter
$K_{0}$ (discussed in Sect.~\ref{FFModel}).~Furthermore, one has to remark that the solar wind usually takes of the order of or more than
one year to reach the border of the heliosphere.~As a consequence, the above parameters are
locally evaluated within the heliosphere, allowing the modulation treatment to better
(or \textit{dynamically}) account for the effects of solar activity as a function of the distance
from the Sun.~In addition, the current treatment accounts for  effects due to the charge sign of particles (i.e., the so-called
\textit{particle drift effect}),~e.g., those related, for instance, to a) the curvature and gradient
of the interplanetary magnetic field (IMF) and b) the extension of the neutral current sheet inside
the heliosphere.~Thus the model introduces a dependence on the sign solar-field polarity ($A$)~\citep[e.g., see][]{Clem,Drift}.~The
present code allows the fluxes of protons (as well antiprotons) and helium nuclei to be modulated from the border of the heliosphere down to Earth - but outside Earth's magnetosphere~\citep{mi_jgr} -
in order to compare them with the available experimental observations.~Furthermore, electrons and positrons modulated spectra can be derived accounting for the additional collision, radiative and inverse Compton energy-losses \citep[see][]{mauroPoster2010}.~
\par
In the next
sections, the heliosphere, drifts, diffusion tensor, determination of the diffusion parameter,
dependence of both the solar wind and IMF on the radial distance and helio-colatitude,
neutral current sheet are discussed (Sects.~\ref{Model}--\ref{antysymmetric_part}).~Then, the implementation of the mathematical model and the parametrization with the dynamical
treatment of heliosphere are treated (Sects.~\ref{dynamic},~\ref{Code}).~Finally, comparisons among obtained modulated spectra of differential intensities with
those experimentally observed are performed and discussed (Sects.~\ref{Results}--\ref{Heliosph_latid_dep}).
\section{Heliosphere and Drift Mechanisms}
\label{Model}
The transport of galactic protons (GP) inside the heliosphere was initially treated
by~\citet{parker65}, who demonstrated that - in the framework of statistical physics -
the random walk of the cosmic ray particles is a Markoff process, describable by a
Fokker--Planck equation (hereafter FPE)
(e.g., see also~\citealt{axford,Fisk76,potgieter93},~and also~Sections~4.1.2.4 of~\citealt[][and references therein]{RancBook}).~Thus (at the time $t$),
the number density\footnote{The equivalent expression in terms of the omnidirectional distribution function of CR particles with momentum $\vec{p}$,
at the position $\vec{r}$ and time $t$ can be found, for instance, expressed in~Equation (1) of~\citet{Potgieter_1998} (see also references therein).}
$U$ of GPs per unit
interval of particle kinetic energy $T$ (the so-called differential density)
can be obtained from the solution of the FPE:
\begin{equation}\label{eq_parker}
    \frac{\partial U}{\partial t} =\frac{\partial}{\partial
x_i}\left( K^S_{ij} \, \frac{\partial U}{\partial x_j} \right) + \frac{1}{3}\,\frac{\partial V_{{\rm sw},i}}{\partial
x_i}\,\frac{\partial}{\partial T}\left(\alpha_{\rm rel} T\, U\right)-\frac{\partial}{\partial x_i}\left(V_{{\rm sw},i}\,U\right)
-\frac{\partial}{\partial x_i}\left( v_{d,i}  U\right)
\end{equation}
(e.g.,~see \citealt{Jokipii77}, Equation~(4.75) in Section~4.1.2.6 of~\citealt{RancBook} and references therein) with $V_{{\rm sw},i}$ the solar wind velocity
along the axis $x_i$,
\begin{equation}\label{drift_antisym_ten}
     v_{d,i}  = \frac{\partial K^A_{ij}}{\partial x_j}
\end{equation}
the drift
velocity (e.g., see~\citealt{Jokipii77,jokipii77b} and also~\citealt{art_ste} and references therein), $K^A_{ij}$ and $K^S_{ij}$ the antisymmetric and symmetric part of the diffusion tensor - respectively -,
\[
\alpha_{\rm rel}=\frac{T+2 m_r c^2}{T+m_r c^2}
\]
and
$m_r$ the rest mass of the proton.~The number density $U$ is related to the differential intensity $J$ as:
\begin{equation}\label{eq::J}
 J=\frac{v\,U}{4\pi},
\end{equation}
where $v$ is the speed of the GCR particle.~Equation~(\ref{eq_parker}) - as well known - describes
i) the diffusion of GCRs by magnetic irregularities, ii) the so-called \textit{adiabatic-energy changes}
associated with expansions and compressions of cosmic radiation, iii) the convection effect resulting from
the solar wind with velocity $\vec{V}_{{\rm sw}}$ and iv) the drift effects related to the \textit{drift velocity}
($\vec{v}_{d}$).~In turn, the drift velocity is determined by the antisymmetric part of the diffusion
tensor [see Eq.~(\ref{drift_antisym_ten}) and Sect.~\ref{antysymmetric_part}] which
accounts for gradient, curvature and current sheet drifts of particles in the IMF,~i.e.,
it depends on the charge sign of particles.~
\par
Furthermore - as discussed by \citet{jokipii77b} -,~one can re-write Eq.(\ref{eq_parker}) as
\begin{equation}\label{eq_parker1}
    \frac{\partial U}{\partial t} =\frac{\partial}{\partial
x_i}\left( K^S_{ij} \, \frac{\partial U}{\partial x_j} \right) + \frac{1}{3}\,\frac{\partial V_{{\rm sw},i}}{\partial
x_i}\,\frac{\partial}{\partial T}\left(\alpha_{\rm rel} T\, U\right)-\frac{\partial}{\partial x_i}\left[\left(V_{{\rm sw},i} + v_{d,i} \right)\,U\right].
\end{equation}
Thus, one obtains that drift effects are accounted for by a convection velocity in which the drift velocity is
added to the solar wind velocity.~In this way, the resulting \textit{effective
convection velocity} may non-negligibly differ from that due to the solar wind; but -
as remarked by \citet{jokipii77b} - noting that $\nabla \cdot \vec{v}_{d} = 0$, one finds that drift effects
do not contribute to the
adiabatic-energy changes [second right-hand term of~Eqs.~(\ref{eq_parker},~\ref{eq_parker1})].~Even if drift
effects are included\footnote{One can see the discussions in~\citep{parker65,Jokipii_1970,jokipii77b,Jokipii77,Potgieter_1998}.} in Eqs.~(\ref{eq_parker},~\ref{eq_parker1}),
some modulation models\footnote{Like, for instance, the so-called \textit{force-field model} (FFM)~\citep[see][]{gleeson1968}.
} neglected it~\citep[e.g., see][and references therein]{Jokipii77,ilya05}.~Gradients of particle
density can also result from the convection effect.~Drift mechanisms can modify both the radial and (solar)
latitude dependence of the gradient magnitude.~For instance,
drift motions can affect modulated GCR spectra by redirecting
particles within the heliosphere \citep{Jokipii77}.~When the particle Larmor radius is much
shorter than the magnetic-field scale length, drift effects can be taken into account by
evaluating the average distance in which a relevant field variation occurs.~Drift
effects affect particle motions over large distances due to the large scale variation of
the IMF strength.~Different intensities of GCR modulation were observed in time periods with opposite
field polarity, for instance, by \citet{Emerson,Garcia-Munoz,Clem1,Drift}.~Thus, it is necessary to
explicitly consider particle drifts inside the equation of
propagation of GCR.~
\par
As well known for a reference system with the 3rd coordinate along the average magnetic field, the symmetric part of the diffusion tensor (or coefficient) - for an isotropic perpendicular diffusion - includes both the transverse ($K_{\perp}$) and parallel ($K_{||} $) components~\citep[e.g., see][]{Jokipii1971,Potgieter85,potgieter94}.~In turn, for a standard Parker field [Eq.~(\ref{FUN_ParkerField})] these two components are related to the radial component in heliocentric spherical coordinates as
\begin{equation}\label{radial_comp}
    K_{rr} =
K_{||} cos^{2}\psi + K_{\perp} sin^{2}\psi,
\end{equation}
with $\psi$  the
angle between radial and magnetic field directions - the so-called \textit{spiral angle}
[Eq.~(\ref{spiral_angle})] - \citep[e.g., see][]{Fisk76,potgieter94}
and $K_{\theta\theta} = K_{\perp}$,
where $\theta$ is the polar angle \citep{potgieter93}.~It has to be remarked that the general
transformations of the symmetric and antisymmetric parts of the diffusion tensor from field-aligned
to heliospheric (spherical) coordinates can be found in~\citep{burg2008}.~Furthermore, it has to be remarked that a general discussion about the role of parallel and perpendicular diffusion is available in~\citealt{Giacalone}.
\par
\citet*{potgieter94}~\citep[see also][]{potgieter93} suggested that the parallel diffusion coefficient is given by
\begin{equation}\label{parallel_comp}
K_{||} \approx \beta \, k_1(r,t) \,K_{P}(P,t)\,\left[\frac{B_{\oplus}}{3B}  \right]
\end{equation}
with $\beta = v/c$ , $v$ the particle velocity and $c$ the speed of light; the diffusion parameter
$k_1$ accounts for the
dependence on the solar activity and is treated in Sect.~\ref{FFModel}; $ B_{\oplus}$ (typically $ \approx 5$\,nT)
is the value of IMF at Earth's orbit but it varies as a function of the time;
$B$ is the magnitude of the large scale IMF (discussed in Sect.~\ref{HMF}), thus, it
depends on the heliospheric region (Sect.~\ref{dynamic}) through which GCRs are transported; finally, the term
$K_{P}$ takes into account the dependence on the rigidity $P$ of the GCR particle and is usually expressed
in GV.~To a first approximation, one can assume that
\begin{equation}\label{KP_Linear}
    K_{P} \approx P
\end{equation}
for particle rigidities above a threshold value $P_{\rm th}$ within the rigidity range (0.4--1.015)\,GV, as commonly supposed by many authors~\citep[e.g., see][]{gloek66,gleeson1968,perko1987,potgieter94,strauss2011}.~In the present model, $K_P$ is assumed to be equal to the value of the rigidity ($P$) above the upper limit of the $P_{\rm th}$ range, i.e., for proton kinetic
energies $\gtrsim 0.444\,$GeV (see Sects.~\ref{H_A_Results},~\ref{R_L_Tilt_Angle}).~Below $P_{\rm th}$, it can be usually approximated to a constant~\citep[e.g., see][]{perko1987,potgieter94,wibber2001,strauss2011}.~It has to be remarked that nowadays treatments resulting in a more complex dependence of the diffusion tensor on rigidity are proposed by several authors \citep[e.g., see][and references therein]{ferreira2001,pei_diffTensor}.~Some of these studies are motivated from dealing with magnetohydrodynamic turbulence in the expanding solar wind and/or accounting observations carried out on data of
low energy electrons collected using spacecrafts [for instance, (3--10)\,MeV from Ulysses spacecraft in~\citep{ferreira2001} and 16\,MeV from Pioneer 10 in~\citep{potgieter2002}].
%,~i.e., the lower energy value of the simulated differential intensities discussed
%in Sect.~\ref{antysymmetric_part}.~
\par
In heliocentric spherical coordinates, the perpendicular diffusion coefficient has two components,
one along the radial direction, $K_{\perp r}$, the other one for the polar direction $K_{\perp \theta}$.~
$\rho_k$ is
%$(K_{\perp})_{0}$,
the ratio between perpendicular (in the radial direction) and parallel diffusion coefficients,~i.e., $
K_{\perp r} = \rho_kK_{||}$.~In the present model, we use $\rho_k = 0.05$: this value is in the mid of the range suggested by \citet*{palmer}
(see also \citealt{Giacalone_1988} and Section~6.3~of \citealt{burger2000}).~The value of the perpendicular diffusion
coefficient in the polar direction ($K_{\perp \theta}$) can be assumed to be almost equal to that radial
($K_{\perp r}$)~\citep[e.g., see][and references therein]{potgieter2000}.~However,~\citet*{potgieter2000} suggested the usage of an \textit{enhanced} $K_{\perp \theta}$ in the polar regions
in order to reproduce the
amplitude and rigidity dependence of the latitudinal gradients of GCR differential intensities for
protons and electrons~\citep[e.g., see][]{potgieter97,Ulysses}.~He introduced a sharp transition (via a transition function, e.g.,~see Figure~7 in that article) in the colatitude regions $120^{\circ}  \lesssim \theta \lesssim 130^{\circ}$ and $60^{\circ} \gtrsim \theta \gtrsim 50^{\circ}$.~He also derived that $K_{\perp \theta}$ has to be increased by a factor of about (or larger than) 10;~\citet*{FER_POT} used a factor of 8.~In the current code, $K_{\perp \theta}$ is given by:
\begin{eqnarray}\label{enhaced_K}
K_{\perp \theta}  = \left\{
\begin{array}{lcc}
10\, K_{\perp r}, %& %\qquad
%\qquad\qquad &
\textrm{ in the polar
regions},\\
K_{\perp r}, %& \qquad \qquad
\textrm{ in the
equatorial region},
\end{array}\right.
\end{eqnarray}
where the polar regions correspond to colatitudes with $\theta \lesssim 30^{\circ}$ or $\theta \gtrsim 150^{\circ}$, while the equatorial region corresponds to colatitudes with $30^{\circ}  \lesssim \theta \lesssim 150^{\circ}$.~The solar colatitudes of $30^{\circ}$ and $150^{\circ}$ correspond to those at which the SW speed becomes constant in periods not dominated by high solar activity [Eq.~(\ref{FUN_swfunct1})].~The usage of the transition function can be fully implemented in current treatment, but is not required with the present overall code accuracy.~In fact, the results obtained from the so-called ``L'' \textit{model} (i.e.,~the one with a better agreement with data, see Sects.~\ref{H_A_Results},~\ref{R_L_Tilt_Angle}) indicate that only in periods not dominated by high solar activity the enhancement of $K_{\perp \theta}$ [Eq.~(\ref{enhaced_K})] slightly improve the overall agreement with data by a few percent.~Finally, in Appendix~\ref{app1} the diffusion coefficients in heliocentric polar coordinates are expressed in terms of those parallel and perpendicular to the IMF.
\par
Moreover, it has to be remarked that the diffusion tensor i) is not well determined during solar maxima and
ii) can be adapted to better account for the complex structure of the IMF - which depends on the solar activity -
found with Ulysses spacecraft~\citep[e.g., see][and references therein]{burg2008}.~For instance, Potgieter, Burger and Ferreira~(2001) - see also references therein -
discussed the so-called \textit{propagating diffusion barriers} and suggested a time dependent model for the
diffusion coefficients.~The latter are supposed to be $\propto [B_0/B(t)]^n$, where $B(t)$
is the IMF magnitude at the time $t$ and $B_0=5$\,nT is the average IMF magnitude during minimum modulation
conditions at Earth \citep{potgieter2001b,ferreira2003} and $n$ is the ratio between the actual tilt-angle value
(Sect.~\ref{HMF}) and that close to solar minimum ($7^{\circ}$--$15^{\circ}$)~\citep[e.g., see][]{potgieter2001b,potgieter2001}.~However,
 in the current model the time dependence of diffusion coefficients is taken into account using a diffusion
parameter, which is treated in Sect.~\ref{FFModel}.~The agreement with data
obtained during high solar activity is discussed in Sect.~\ref{H_A_Results}.
\subsection{Diffusion Parameter in the Framework of the Force Field Model}
\label{FFModel}
In the FFM (e.g., see \citealt{gleeson1968,G_U_1971} and also Section~4.1.2.4 of~\citealt{RancBook}), Gleeson \& Axford (1968)
assumed that, at the time $t$, i) modulation effects can be expressed with a spherically symmetric modulated differential
number density $U$ of GCRs, ii) the diffusion coefficient reduces to a scalar\footnote{While in Eq.~(\ref{eq_parker}), it is expressed
by a tensor with a symmetric and an antisymmetric part (see discussion in Sect.~\ref{Model}).} given
by a separable function of $r$ (the radial distance from the Sun) and $P$ (the particle rigidity in GV):
\begin{equation}\label{diffusion_coeff}
   \mathcal{K}(r,t) = \beta k_1(r,t) K_{P}(P,t)
\end{equation}
with $ K_{P}$ from Eq.~(\ref{KP_Linear}) for particle rigidities above $\approx 1\,$GV, and iii) the modulation occurs
in a steady-state condition, i.e., the relaxation time of the distribution is short with respect to the solar cycle duration
so that one can assume that the partial derivative of $U$ with respect to time is zero.~They derived that the differential
intensity [Eq.~(\ref{eq::J})] at a radial distance $r$ is given by the expression
\begin{equation}\label{force_field_solution}
   J (r,E_{\rm t},t) =  J ({r_{\rm tm}},E_{\rm t}+ \Phi_{\rm p})
   \left[ \frac{ E_{\rm t}^2 -  m_{\rm r}^2 c^4} {( E_{\rm t} + \Phi_{\rm p} )^2-  m_{\rm r}^2 c^4} \right],
\end{equation}
where $J ({r_{\rm tm}},E_{\rm t}+ \Phi_{\rm p})$ is the undisturbed
intensity beyond the solar wind termination located at a radial distance $r_{\rm tm}$ from the Sun; $E_{\rm t}$
is the total energy of the particle with rest mass $m_{\rm r}$ and, finally, $\Phi_{\rm p} $ is the so-called
force-field energy loss \citep{gleeson1968,G_U_1971}.~When modulation is small
[i.e., for $\Phi_{\rm p} \ll m_{\rm r}c^2, T$]~\citep{gleeson1968,G_U_1971,G_U_1973}, they determined that
\[
\Phi_{\rm p}  = \frac{Z e P}{K_{P}(P,t)}\, \phi_{\rm s}(r,t) \approx Z e \,\phi_{\rm s}(r,t),
\]
where $Z e $ is the particle charge and $\phi_{\rm s}(r,t)$ is the so-called \textit{modulation strength}
(or \textit{modulation parameter}) usually expressed in units of GV (or MV).~Assuming that $V_{\rm sw}$ (the solar wind speed) and $k_1$ are almost
constant, $\phi_{\rm s}(r,t)$ is linearly dependent
 on $ \left(r_{\rm tm} -r \right)$ \citep[e.g., see Equation (4.64) of][]{RancBook},
%\begin{equation}\label{md_par_red}
%     \phi_{\rm s}(r,t) \approx \frac{V_{\rm sw}(t) \left( r_{\rm tm} -r  \right)}{3
%    k_1(t)} ,
%\end{equation}
from which one gets that the diffusion parameter is given by
\begin{equation}\label{md_par_red2}
   k_1(t)   \approx \frac{V_{\rm sw}(t) \left( r_{\rm tm} -r  \right)}{3 \phi_{\rm s}(r,t)
    } ,
\end{equation}
i.e., $k_1$ [similarly to $\phi_{\rm s}(r,t)$] is linearly dependent on
$ \left(r_{\rm tm} -r \right)$.~As already mentioned, in the FFM the diffusion coefficient
$\mathcal{K}(r,t)$ is a scalar quantity and does not
account for effects related to the charge sign of the transported
particles.~$ \phi_{\rm s}(r,t)$ is independent of the species of GCR particles
(e.g., see discussion at page 1014 of \citealt{gleeson1968} or Equation (1) of Usoskin and collaborators 2005,
and also \citealt{Bobik_2011,Bobik_2011b}).~Usoskin and collaborators 2005 monthly determined the values of the modulation strengths
[$\phi_{\rm s}(r_{\rm Earth})$] for the time period from 1951 up to 2004
using measurements of neutron monitors (i.e., located at $r_{\rm Earth} = 1\,$AU);
while the values of solar wind speeds are available from NASA/GSFC's OMNI data set through OMNIWeb.~
%%%
\begin{figure}[hbt!]
\centering
\includegraphics[width=4.2in]{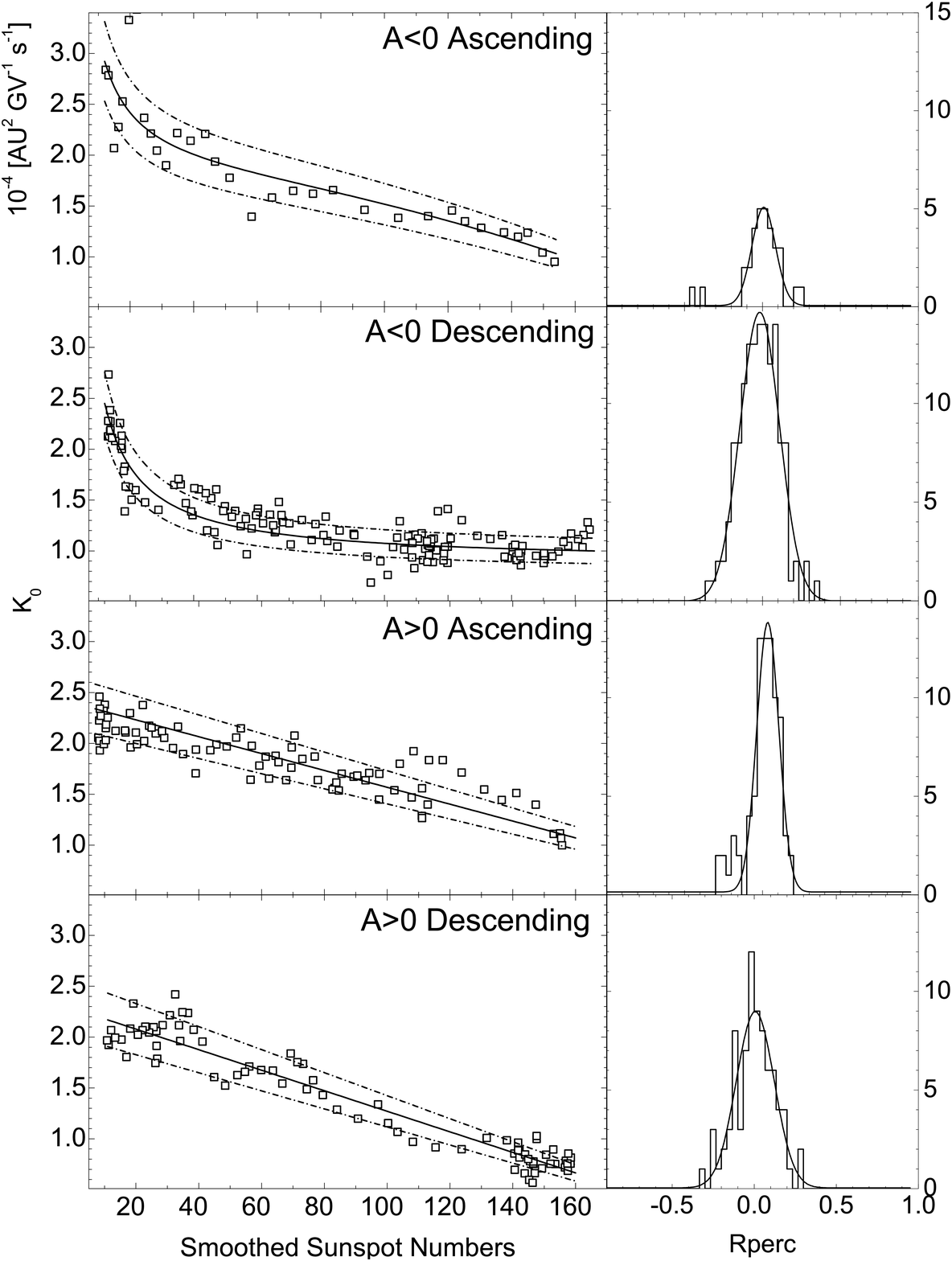}
\caption{Diffusion parameter $K_{0}$ (left side) and percentage differences $R_{\rm perc}$ [Eq.~(\ref{Rperc})] (right side) %- ascending and descending phases for both negative
%and positive solar magnetic-field polarities -
as a function of the SSN value;
the central continuous lines are obtained from a fit of $K_{0}$ with respect to SSN values in the range $10 \lesssim \textrm{SSN} \lesssim 165$;
the dashed and dotted lines are obtained adding (top) or subtracting (bottom) one standard deviation from the fitted values.}
\label{fig_sim1}
\end{figure}
%%%%%%
%%%%%%%%
\begin{table}[t]
\begin{center}
{\renewcommand{\arraystretch}{1.25}\begin{tabular}{|c|c|c|c|c|}
\hline Solar polarity&  $A<0$ & $A<0$ &  $A>0$  &  $A>0$\\
and phase  & ascending   & descending   & ascending    & descending  \\
\hline $c_1$ & $+0.0001686$ & $+8.872\!\times\!10^{-5} $ &  $+2.39708\!\times\!10^{-4}$  & $+2.28037\!\times\!10^{-4}$ \\
\hline $c_2$ & $+0.001488$  & $+0.001874$  & &   \\
\hline $c_3$ &              &  & $-8.28987\!\times\!10^{-7} $ & $-1.00984\!\times\!10^{-6} $ \\
\hline $c_4$ & $-3.164\!\times\!10^{-9}$ &  &  & \\
\hline
\end{tabular}\caption{Parameters of the polynomial expression~(\ref{K0_SSN}).}\label{coefficients_fit}}
\end{center}
\end{table}
%%%%%%%%%%%
\par
To determine $\phi_{\rm s}(r_{\rm Earth})$, Usoskin and collaborators (2005) used an approximated expression of the Local Interstellar Spectrum
(LIS) for protons %(which by far constitute the largest component of GCR)
from Burger, Potgieter and Heber (2000).~In practice,
 their spectrum differs from that due to Burger, Potgieter and Heber (2000) by about or more than 5\% at kinetic energies
lower than about 117\,MeV.~Furthermore, in the present work we found that the error-weighted average
of the differential spectral index, $\gamma_{\rm wa}$ [Eq.(\ref{Gammawa}) and discussion in Sect.~\ref{LIS_discussion}], of the proton LIS is only compatible,
within one standard deviation, with the differential spectral index ($\gamma =2.78$) of the spectrum from Burger and Potgieter (1989)
or Burger, Potgieter and Heber~(2000).~It has to be remarked that the latter spectral index is the one used by Usoskin and collaborators (2005).~Usoskin and collaborators (2005) [see Appendix~A in that article]
also found that using other commonly adopted LIS's their corresponding values of the modulation strengths follow a linear relation
with respect to  $\phi_{\rm s}(r_{\rm Earth})$.~However, the differential spectral indexes of these spectra are not compatible
within three or more standard deviations with that found in Sect.~\ref{LIS_discussion}.~Moreover, it has to be noted that the response of neutron monitors has to be evaluated by combining a) the
effects of both the geomagnetic cutoff rigidity \citep{ilya05} which results in a reduced sensitivity of detection apparata and b)
the so-called \textit{atmospheric yield function} \citep{clem2000}.~Thus, one finds that i) the contribution of the
GCRs with rigidities below 2\,GV amounts to about or less than 1.1\% of the total neutron monitor counts due
to particles with energies up to about 50\,GV and ii) the maximum of neutron monitor sensitivity -~i.e.,
the maximum of the response function [see Figure~7 of~\citet{clem2000}] - occurs in the
rigidity interval (3--15)\,GV.~In addition, Boella and collaborators (2001) determined -
using IMP8 satellite data during the period 1973--1995 - that charge effects (discussed
in Sects.~\ref{Intr},~\ref{Model}) result in a variation of proton or helium fluxes during solar minima
with opposite magnetic field polarities of $14\pm6$\% at $\approx 300\,$MeV/nucleon.~This
variation steadily decreases with increasing energy [e.g., see Figure~4.13 at page~378 of~\cite{RancBook}].~As a consequence, $\phi_{\rm s}(r_{\rm Earth})$ is expected to be marginally
affected by drift effects.~%In Sect.~\ref{antysymmetric_part}, it will be shown that the ratios of modulated
%differential intensities for opposite magnetic field polarities agree with the experimental data from Boella and collaborators (2001)
%[and Figure~4.10 at page~328 of Leroy and Rancoita~(2009)] using both the symmetric -
%with the currently determined diffusion parameter $K_0$ - and the antisymmetric -
%which deals with the drift effects - part of the diffusion tensor.
\par
$k_1$ [Eq.~(\ref{md_par_red2})] depends on the value of solar wind termination
located at a radial distance $r_{\rm tm}$ related, in turn, also to solar wind
speed [e.g.,~see Chapter~7 of~\cite{Meyer-Vernet}, and Sections~4.1.2.3,~4.1.2.4 of~\cite{RancBook}].~In the present
simulation code, the \textit{effective heliosphere} assumes that the solar wind termination is located at 100\,AU
(see a further discussion in Sects.~\ref{dynamic},~\ref{Effective_Hel}).% in order to calculate the
% modulated differential intensities at $r_{\rm Earth}$.
~Therefore, from the diffusion parameter $k_1$
one has to derive that ($K_0$) for an effective heliosphere with a radial extension of 100\,AU.~In practice,
for a radial extension of 100\,AU the diffusion parameter $K_0$ [Eq.~(\ref{md_par_red1}] replaces $ k_1$ in Eq.~(\ref{parallel_comp}) (for instance, see Appendix~\ref{app1}) and in allows one to obtain similar modulation
effects on the differential intensities of GCRs  with respect to those obtained using $k_1$ when the
heliosphere has a variable radial extension $r_{\rm tm}$.~Using Eq.~(\ref{md_par_red2}) one obtains
\begin{equation}\label{md_par_red1}
    K_0  \approx k_1   \frac{99\,\textrm{AU}}{\left( r_{\rm tm} -r_{\rm Earth}  \right)} = 99\,\textrm{AU}\left[\frac{V_{\rm sw} }{3
     \,\phi_{\rm s}(r_{\rm Earth})}\right],
\end{equation}
where 99\,AU (as already mentioned) is the distance of the Earth from the border of the effective heliosphere used in the current simulation code.~In Fig.~\ref{fig_sim1}, the diffusion parameter $K_{0}$ - obtained from Eq.~(\ref{md_par_red1}) - is shown as a function of the corresponding value of smoothed sunspot number, SSN, \citep{ssn}.~The $K_{0}$ data had to be subdivided in four sets, i.e., ascending and descending phases for both negative and positive solar magnetic-field polarities.~For each set, the data could be fitted with a \textit{practical relationship} (see Fig.~\ref{fig_sim1}) between $K_{0}$ and SSN values for $10 \lesssim \textrm{SSN} \lesssim 165$,~i.e., finding
\begin{equation}\label{K0_SSN}
   K_{\rm F} = c_1 +  c_2 \times \textrm{SSN}^{-1} +  c_3 \times \textrm{SSN}  +c_4 \times \textrm{SSN}^2
\end{equation}
with the parameters $c_i$ shown in Table~\ref{coefficients_fit}.~In addition, the data were found to
exhibit a Gaussian distribution of percentage differences ($R_{\rm perc}$) of $K_{0}$ values from the corresponding fitted values $K_{\rm F}$, with
\begin{equation}\label{Rperc}
 R_{\rm perc}=\frac{K_{\rm F}-K_0}{K_{\rm F}}.
\end{equation}
The rms\index{rms} values of the Gaussian distributions were found to be $\approx 0.1339, 0.1254, 0.1040, 0.1213 $
for the phases ascending with $A<0$, descending with $A<0$, ascending with $A>0$, descending with $A>0$,
respectively.~From the practical relationship found [Eq.~(\ref{K0_SSN})], we can use the estimated SSN values
to obtain the diffusion parameter $K_{0}$ at times beyond 2004.~This procedure allows one to extend
the $\approx 40$ years period by exploiting the practical relationship between the fitted $K_0$ values and the SSN values (one of
the main parameters related to the solar activity).~In addition,
we introduced in our code a Gaussian random variation of $K_{0}$
with rms's corresponding to those found for each subset of data.~Results of the simulation with and without the
Gaussian variation are consistent within the uncertainties of the code.~Furthermore, it can be noted that $K_{0}$ results in providing an overall increasing (for $r_{\rm tm}$ lower than 100\,AU) or decreasing (for $r_{\rm tm}$ larger than 100\,AU) of modulation effects.~A tuning of the effective extension of the heliosphere and its dependence on the solar activity is likely to be obtained using the experimental data from long-duration accurate observations, like those from the AMS-02 spectrometer.
\section{Solar Wind and Latitudinal Dependence IMF}
\label{HMF}
Parker~(1958) suggested that the solar corona is stationary expanding due to an outflow of the coronal plasma -
generating the so-called \textit{solar wind} - with a spherically symmetric velocity.~In his model,
the solar wind speed becomes almost constant ($V_{\rm sw}$) beyond a radial distance from the Sun $r_b \approx (0.3$--0.4)\,AU
(e.g.,~see Figure~1 of~\citealt{parker58}).~Furthermore, the magnetic-field lines
are frozen in the streaming particles of which the solar wind consists.~Thus, beyond $r_b$, in a
spherical reference frame rotating with the Sun the components of the outward velocity of a plasma
element carrying the magnetic field are: $V_r=V_{\rm sw}$, $V_\theta=0$ and
$V_\phi=\omega(r-r_b)\sin\theta$ with $\omega$ the angular velocity of the Sun.~The streamline has the shape of an Archimedean spiral (termed \textit{Parker spiral}).
\par
In heliocentric spherical coordinates, the standard Parker
spiral field can be expressed as~\citep[e.g.,~see Equation~(2) of][]{Hattingh95}:
\begin{equation}\label{FUN_ParkerField}
\vec{B}_P =   \frac{A}{r^2}\left(\vec{e}_r - \Gamma\,
\vec{e}_\phi\right)\left[1-2H(\theta-\theta')\right],
\end{equation}
where $A$ is a coefficient that determines the
field polarity and allows $|\vec{B}_P|$ to be equal to $ B_{\oplus}$ (Sect.~\ref{Model}),~i.e.,
the value of IMF at Earth's orbit as extracted from NASA/GSFC's OMNI data set through OMNIWeb~\citep{SW_web}; $\vec{e}_r $ and $\vec{e}_\phi $ are unit
vector components in the radial and azimuthal direction,
respectively; $\theta$ is the co-latitude (polar angle); $\theta'$ is the polar angle determining the position of
the HCS~\citep{Jokipii1981}; $H$ is the Heaviside function, thus,
$[1-2\,H(\theta-\theta')]$ allows $ \vec{B}_P$ to interchange the sign in the two regions
- above and below the heliospheric current sheet (HCS) - of the heliosphere; finally,
\begin{equation}\label{spiral_angle}
\Gamma= \tan\psi =\frac{\omega \,(r-r_b) \sin\theta}{V_{\rm sw}}
\end{equation}
with $\psi$ the spiral angle.~In the present,
model $\omega$ is assumed to be independent of the heliographic latitude and equal to the \textit{sidereal rotation} at
the Sun equator.~However, the simple representation of the Parker spiral [Eqs.~(\ref{FUN_ParkerField},~\ref{spiral_angle})]
based on a constant solar wind speed needs to be complemented with the present knowledge of the speed [$V_{\rm sw}(\theta)$]
dependence on solar colatitude.~Large variations of the solar wind structure were observed for solar latitudes up to $|80^{\circ}|$ by Ulysses spacecraft~\citep{Wenzel1998}.~For
instance, during a period of low solar activity the solar
wind speed increases by almost a factor two from the ecliptic plane
to poles, thus subdividing the heliosphere in two regions with slow and
fast solar wind~\citep{McComas2000}.~For representing the observed speeds, Fichtner, Ranga and Fahr~(1996) suggested
that the solar wind speed may be proportional to $(1+\cos^2 \theta)$.~In the present model we use:
\begin{eqnarray}\label{FUN_swfunct1}
V_{\rm sw}(\theta) = \left\{
\begin{array}{lcc} V_{sw_{\rm max}}, %& %\qquad
%\qquad\qquad &
\textrm{ for }  \theta  \leq 30^{\circ} %\qquad
\textrm{ and }
%\qquad
\theta \geq 150^{\circ} ,\\
V_{sw_{\rm min}}\times ( 1 + \left|\cos \theta\right| ), %& \qquad \qquad
\textrm{ for } %\qquad&
30^{\circ }< \theta < 150^{\circ}
\end{array}\right.
\end{eqnarray}
with $V_{sw_{\rm max}}
\simeq 760$\,km/s~\citep[e.g., see][]{McComas2000}] and $V_{sw_{\rm min}} $
is the corresponding value extracted from NASA/GSFC's OMNI data set through OMNIWeb~\citep{SW_web}.~Equation~(\ref{FUN_swfunct1})
exhibits a slightly better agreement with observed data than that proposed by Fichtner, Ranga and Fahr~(1996).~Jokipii and K\'ota (1995)
and Pommois, Zimbardo and Veltri (2001) proposed other functions for such periods.~However,
these functions depends on an additional parameter related to the latitudinal extension of the
region with a slow solar wind.~The parameter can be determined only using measurements to be performed
largely outside the ecliptic plane, like those due to Ulysses spacecraft.~Thus, Eq.~(\ref{FUN_swfunct1})
has the advantage to allows one to more generally treat periods of low solar activity.~Furthermore, McComas and collaborators (2000)
observed that during the Sun's approach to solar maximum a) the coronal structure becomes increasingly complex
and b) the magnetic field becomes less dipolar.~In the present model, for the solar wind we assume a speed independent of the colatitude in periods characterized by a large solar activity.~As previously, the speed value is
extracted from NASA/GSFC's OMNI data set through OMNIWeb~\citep{SW_web}.
\par
\citet{1989JGR_94.2323P} pointed out how \textit{classical drift} modulation models -
based on the Parker magnetic-field up to the polar region - encounter difficulties (see also Sect.~\ref{Heliosph_latid_dep})
in accounting for the significantly lower latitudinal dependence of CRs intensity.~\citet{simpson1996} subsequently observed this phenomenon using
Ulysses spacecraft data collected in the inner heliosphere.~Heber and collaborators~(1998) remarked that a) one needs to assume an anisotropy of perpendicular diffusion
coefficient and enhancement in the latitude direction (as already treated in Sect.~\ref{Model}), and b) Parker's IMF has to be modified\footnote{Limited to polar regions, Fisk~(1996) proposed that magnetic-field lines are non radially expanding.~In addition,
Hitge and Burger~(2010) have attempted to merge Parker and Fisk magnetic-fields into a hybrid field.} as proposed by Jokipii and K\'ota~(1989).~
%%%%%%%%%%%%%%%%%%%%%%%%%%%%%%%%%%%%%%%%%%%%%%%%%%%%%%%%%%%%%%%%%%%%%%%%%%%%%%%%%%%%%%%
\begin{figure}[t]
\begin{center}
 \includegraphics[width=0.7\textwidth]{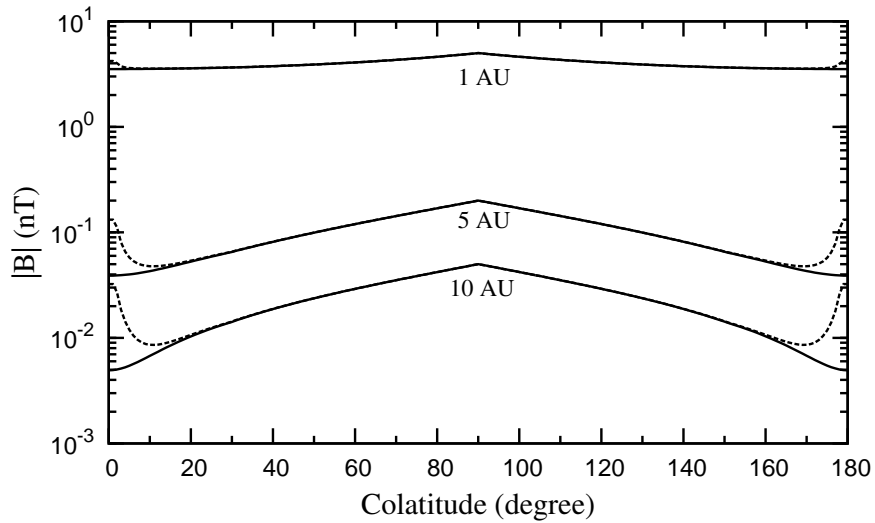}
 \caption{Magnitude ($|B|$) of the IMF (dashed line) from \citep{JokipiiKota89} - computed using Eqs.~(\ref{lat_comp}--\ref{B_Kota}) - compared with that
  from Parker (solid line) [Eq.~(\ref{FUN_ParkerField})] at 1, 5 and 10\,AU as a function of the colatitude.~For the purpose of this calculation, at 1\,AU and $90^\circ$ $|B| = 5\,$nT.}\label{IMG_MagB3D}
\end{center}
\end{figure}
%%%%%%%%%%%%%%%%%%%%%%%%%%%%%%%%%%%%%%%%%%%%%%%%%%%%%%%%%%%%%%%%%%%%%%%%%%%%%%%
\par
In the present model, the magnitude of the magnetic field [Eq.~(\ref{FUN_ParkerField})] is enhanced introducing a
small latitudinal component~\citep[e.g.,~see][]{langner2004,2004JGRA_10901103L}
\begin{equation}\label{lat_comp}
    B_{\theta} = \frac{A}{r r_{\odot}}\delta(\theta)
\end{equation}
with $r_{\odot}$ the solar radius,
\begin{equation}\label{delta_sin}
    \delta(\theta)=\frac{8.7\times 10^{-5}}{\sin \theta};
\end{equation}
for $\theta \lesssim 1.7^{\circ}$ and $\theta \gtrsim 178.3^{\circ} $, $\delta$ is $\simeq 3\times 10^{-3}$ \citep{Fichtner96}.~It
has to be noted that
Eqs.~(\ref{lat_comp},~\ref{delta_sin}) allows one to obtain $\nabla \cdot \vec{B} =0$.~The
magnitude of the magnetic field used in the current model is given by \citep{jokipiikota1989}:
\begin{equation}\label{B_Kota}
  B = \frac{A}{r^2}\sqrt{1+\Gamma^{2}+\left(\frac{r}{r_{\odot}}\right)^{2}\delta^{2}(\theta)}.
\end{equation}
In Fig.~\ref{IMG_MagB3D}, the magnitude ($|B|$) of the IMF from \citep{JokipiiKota89}
is compared with that
  from Parker [Eq.~(\ref{FUN_ParkerField})] at 1, 5 and 10\,AU as a function of the colatitude: the field magnitude significantly increases
  in the polar regions (colatitude $\lesssim
  10^\circ$ and $\gtrsim 170^\circ$), while it is almost unchanged
  in the ecliptic region (colatitude $\approx 90^\circ$).~$|B|$ (Fig.~\ref{IMG_MagB3D}) was computed using Eqs.~(\ref{lat_comp}--\ref{B_Kota}).~As discussed by Haasbroek and Potgieter~(1995), the above modification of the Parker IMF allows the modulation effect in the polar regions to be increased and, subsequently, more realistic radial and latitudinal gradients.
\par
As well known~\citep[e.g., see][and references therein]{Bravo}, during several years around solar minimum
the general structure of the solar magnetic-field
is more or less axially symmetric, dominated by the dipole component.~These periods are characterized
with corresponding low values of the tilt angle ($\alpha_{\rm t} < 30^\circ$,~\citealt{potgieter2001}).~As the solar activity increases,
the dipolar structure inclines more and more with respect to the rotation axis and
the effect of higher multipoles becomes more relevant~\citep{sanderson2003}.~During the years of high activity, the structure
of the solar magnetosphere is very complex and the dipole component is very tilted~\citep[e.g., see][]{sanderson2003,wang2002}.~These periods are characterized
with corresponding large values of the tilt angle ($\alpha_{\rm t} > 75^\circ$,~\citealt{potgieter2001}).~Finally,
one can remark that, dealing with neutron-monitor measurements,~\citet{cliver2001} concluded that a diffusion/convection dominated modulation occurs when the tilt angle
exceeds $50^\circ$ [for a previous discussion and illustration with numerical models,~e.g., see~\citet{1995Ap&SS.230_393P} and references therein].
\par
In the
current model the evolution of the solar magnetosphere and, subsequently, of the IMF is (partially)
taken into account using the diffusion parameter (treated in Sect.~\ref{FFModel}) and the actual value of
tilt angle.~In fact, the diffusion parameter depends on the solar phase (ascending and descending) and
solar polarity (positive and negative), and is practically related to the actual value of smoothed sunspot number via Eq.~(\ref{K0_SSN}); while the tilt angle
allows one to gradually modify the contribution of drift effects to modulation (Sect.~\ref{antysymmetric_part}).~The agreement
with data obtained during high (low) solar activity is discussed in Sect.~\ref{H_A_Results} (Sect.~\ref{R_L_Tilt_Angle}).~
\section{The Neutral Sheet and Large Scale Gradients of IMF}
\label{antysymmetric_part}
${K}_{A}$ expresses the value of the \textit{antisymmetric part of the
diffusion tensor} and results from the effects on the motion of cosmic-ray particles due to drift mechanisms.~In a coordinate system with the 3rd
coordinate along the average %magnetic-field -  where $\mathbf{e}_B$ is a unit vector along the average
IMF, one finds~\citep[e.g., see][]{Potgieter85,burghatt95}
\begin{equation}\label{antysmm_ten}
    {K}_{A} = \frac{p v}{3\,Ze |B|} ,
\end{equation}
where $p$, $v$ and $Ze$ are the momentum, velocity and charge of the cosmic-ray particle,
respectively.~Thus, the antisymmetric elements of the diffusion-tensor matrix (Sect.~\ref{Model}) are
\[
K^A_{ij} = {K}_{A} \, \epsilon_{i,j,k} \,\frac{B_k}{|B|}
\]
with $\epsilon_{i,j,k}$ the \textit{Levi-Civita symbol}~\citep[e.g.,~see Equation~(10) of][]{parker65}.
\par
As already mentioned in Sect.~\ref{Model}, the \textit{drift velocity} $\vec{v}_{d}$ [Eq.~(\ref{drift_antisym_ten})]
accounts i) for effects due to gradient and curvature drifts experienced by cosmic-rays particles transported trough the IMF,
ii) net drift effects occurring close to the HCS, where the IMF changes polarity~\citep[e.g., see][]{parker57,burg1985,Potgieter85}
and iii) can be calculated using the antisymmetric part of the diffusion tensor~\cite[e.g., see][and references therein]{parker65,Jokipii77,Potgieter85,burghatt95}.~
\par
Burger, Potgieter and Heber~(2000) [see also references therein and~\citep{palmer,Lock1992}] remarked that the observational
results (carried out below 5\,GV) are consistent with a small (or very small) ratio of the perpendicular to parallel diffusion
coefficients.~As discussed by Parker~(1965), a small value of that ratio indicates that cosmic-rays particles are practically
moving through several gyro-orbits between each scattering event,~i.e., drift motion is \textit{weakly affected by scattering}.~In addition,
for cosmic-rays particles with rigidities $\lesssim (10$--15)\,GV and an IMF expressed by Eqs.~(\ref{FUN_ParkerField},~\ref{B_Kota}),
the particle gyro-radius is smaller (or much smaller) than any local (i.e.,~inside the heliosphere) scale variation of
magnetic field $L\equiv |(1/B) (\partial B_i/\partial x_i)|^{-1}$.~In this way, for regions outside that of HCS,
Isenberg and Jokipii~(1979) remarked that $\vec{v}_{d}$ is determined by the terms due to the gradient and curvature drifts~\citep[e.g., see also][]{parker57,armstrong}.
\par
Potgieter and Moraal~(1985) treated the modulation of GCR's for steady state conditions with
relevant drift effects including that due to a wavy HCS (WHCS).~They succeeded in formulating a 2-D description
(of the WHCS), which - as discussed by Burger and Hattingh~(1995) - is equivalent to the treatment
of transport in a three-dimensional heliosphere with the assumption of an axis-symmetric particle distribution.~Thus,
they allowed one to neglect the azimuthal dependence.~The effect of a WHCS was included via an appropriate modification
of the antisymmetric part of the diffusion tensor.~In this 2-D modeling, the WHCS is described as a wide region whose
width depends on the rigidity of cosmic-ray particles and actual value of the tilt angle ($\alpha_{\rm t}$).~The resulting
drift velocity in heliocentric polar coordinates - as used in the current model - is given by~\citep[e.g., see~Equation~(6) of][]{burghatt95}:
\begin{eqnarray}
% \nonumber to remove numbering (before each equation)
\label{v_drift} \vec{v}_d
 &=&f(\theta)\nabla \times \left( K_A \frac{\vec{B}}{B}\right) +
 \left(\!\frac{\partial f(\theta)}{\partial \theta}\!\right) \frac{K_A}{r}\,\vec{e}_\theta
 \times \left( \frac{\vec{B}}{B}\right) \\
\label{v_drift0}  &=& \vec{v}_{\rm dr} + \vec{v}_{\rm HCS} ,
\end{eqnarray}
where $K_A $ is from Eq.~(\ref{antysmm_ten}), $\theta$ is the colatitude, $f(\theta)$ is a \textit{transition function} that accounts for the effects of a wavy neutral sheet~\citep{Potgieter85}
and $\vec{e}_\theta$ is the unit vector along the latitudinal direction.~$f(\theta)$ is expressed as~\citep[e.g.,~see Equation~(14) of][]{Potgieter85}:
\begin{eqnarray}\nonumber
f(\theta) = \left\{ \begin{array}{ll} (1/a_h)\,\arctan \left\{
\left\{ 1-[(2\,\theta)/\pi] \right\} \tan (a_h)\right\}, \textrm{ if }  c_h<\frac{\pi}{2}, \\
\\ 1-2\,H\left[\theta-(\pi/2)\right], \textrm{ if } c_h=\frac{\pi}{2}
\end{array} \right.
\end{eqnarray}
with $H$ the Heaviside function,
\[
a_h = \arccos \left( \frac{\pi}{2\,c_h} - 1 \right)
\]
~\citep[e.g.,~see Equation~(15) of][]{Potgieter85},
\[
c_h=\frac{\pi}{2}- \frac{1}{2}\,\sin(\alpha_{\rm t} + \Delta \theta_{\rm {HCS}})
\]
~\citep[e.g.,~see Equation~(23) of][]{burger1989},
\[
\Delta \theta_{\rm {HCS}}=\frac{2\,r_p}{r}
\]
($r_p$ is the particle gyro-radius, e.g., see~\citealt{Hattingh95} and also Section~4.2 of~\citealt{burghatt95}), finally,
$f(c_h)= 0.5$ and $f(\pi/2)= 0$.~$\Delta \theta_{\rm {HCS}}$ is determined from the maximum distance that a particle
drifting along the neutral sheet can be away from this sheet~\citep{burger1989}.
The first term ($ \vec{v}_{\rm dr}$) of Eq.~(\ref{v_drift}) accounts for the
gradient and curvature drifts, the second ($\vec{v}_{\rm HCS}$) for drift in the region affected by a
WHCS.~The transition function sets the rate at which the first term of Eq.~(\ref{v_drift}) goes to 0 on
the ecliptic plane ($\theta= \pi/2$)~\citep{Potgieter85}.
%\par
%The agreement of present model with drift effects - observed by Boella and collaborators (2001)
% using the IMP-8 satellite at solar minima on the ecliptic plane - are discussed in Sect.~\ref{IMP8}.
%
\section{Parameters of the Effective Heliosphere used in the Current Model}
\label{dynamic}
As discussed by Potgieter~(2008) (see also references therein), until recently the heliosphere was assumed
to be spherical in most modulation models with an outer boundary at radial distances beyond $\approx 100\,$AU.~Presently,
the heliospheric structure is considered latitudinally asymmetric (particularly) during solar minimum conditions mostly
because the SW depends on the latitude and solar activity (Sect.~\ref{HMF}).~As a consequence,
the position of termination shock (where the SW ram pressure is balanced by interstellar pressure), TS,
can exhibit a latitudinal asymmetry.
\par
Using solar wind speeds observed from Ulysses, Whang and collaborators \citep[e.g., see][]{whang2000,whang2003,whang2004} could estimate
the radial position of TS on and outside the ecliptic plane.~They found that a) on the ecliptic the radial distance of TS is about
 of $ 80\,$AU on average (without
large variation between low  and high solar activities), b) near the ecliptic the radial distance varies by less then 20\,AU and c)
outside the ecliptic plane (e.g., at a latitude of 35$^\circ$) the location of the TS increases by more than or about 50\,AU~\citep{whang2003}.~In addition,
Whang and collaborators estimated that the averaged
value over a 26-years period of the radial distance of the TS increases with latitude [see Table~2 of~\citep{whang2003}].~It is worthwhile
to remark that $\approx 100\,$AU is the averaged value over the corresponding solid angle of the TS location,
which can be obtained from Table~2 of~\citep{whang2003}.~Furthermore~\citep[e.g.,~see][]{Stone2005,Stone2008},
Voyager 1 and 2 reached the TS in 2004 and 2007 located at about 94.0\,AU and 83.7\,AU, respectively,
in agreement with the predictions from Whang and collaborators.~Langner and Potgieter~(2005) treated symmetric
and asymmetric TS models and concluded that for $A>0$ cycle for solar minimum no significant difference occurs;
for $A<0$ cycle differences remain insignificant in nose direction while, approaching the tail direction,
some differences can be appreciated at proton energies below (1--1.5)\,GeV.~However,
 Langner and Potgieter~(2005) and Potgieter~(2008) suggested that, in general,
a symmetric TS with a radial distance of $\approx 100\,$AU is still a reasonable assumption.
\par
In the present model (as already discussed in Sect.~\ref{FFModel}), the effect of the modulation is obtained
for the GCRs propagation trough a symmetric effective heliosphere with a radius of $100\,$AU.
The diffusion parameter $K_0$ %- which depends on ascending and descending phases for both negative and
%positive solar magnetic-field polarities -
is determined (following the procedure described in Sect.~\ref{FFModel})
using the values of modulation strength, SSN values~\citep{ssn} and radius of the effective heliosphere.~Furthermore, it has to be remarked that
(see discussion in Sect.~\ref{FFModel}) the atmospheric yield function results in a diffusion parameter
related to modulated intensities of GCRs (mostly protons) with rigidities above 2\,GV.~%In Sect.~\ref{Effective_Hel},
%one can see that modulated intensities of cosmic-ray protons above 2\,GV are almost independent of varying the effective
%heliosphere boundary when a corresponding re-scaled $K_0$ parameter is taken into account.~
\par
Other parameters (which depend on the solar activity) are the tilt angle $\alpha_{\rm t}$ of the HCS,
magnetic field polarity [related to the sign of the coefficient $A$ in Eq.~(\ref{FUN_ParkerField})], magnetic field magnitude
($B_\oplus$) and solar wind velocity ($V_{\rm sw}$).~The latter two parameters are measured at Earth's orbit.~The polarity of
 the magnetic field and $B_\oplus$ determines the IMF described by means of Eqs.~(\ref{FUN_ParkerField},~\ref{spiral_angle},~\ref{lat_comp}--\ref{B_Kota}).~$\alpha_{\rm t}$ and
the field polarity are used to deal with the drift velocity (as discussed in Sect.~\ref{antysymmetric_part}), which modifies
the overall convection velocity [Eq.~(\ref{eq_parker1})].~Drift contribution is relevant during low solar activity -
e.g., for $\alpha_{\rm t} <30^\circ$ (Sect.~\ref{HMF}) - and decreases with increasing solar activity.~$\alpha_{\rm t}$ values are obtained from Wilcox Solar Observatory~\citep{Hoek95,wsoWeb} and are
calculated using two different models called ``R'' and ``L''.~\citet{2003AdSpR_32_657F,FER_POT}
suggested that ``R'' model accounts for GCR observations during periods of increasing solar activity
(for instance, 1987.4--1990.0 and 1995.5--2000.0), while ``L'' model accounts for periods of decreasing solar
activity (for instance, 1990.0--1995.5 and 2000.0--2010.0).~The implementation of ``R'' and ``L'' models in the current
code is further treated in Sects.~\ref{H_A_Results}--\ref{R_L_Tilt_Angle}.~Finally, the latitudinal dependence (e.g.,~see Sect.~\ref{HMF}) of
the solar wind [Eq.~(\ref{FUN_swfunct1})] depends (at low solar activity) on the values (averaged over 27 days)
of SW speed and on the ecliptic at Earth orbit.
\par
The time spend by the SW
to cover the distance from the outer corona up to the boundary of the effective heliosphere
can be expressed in units of the time needed for a sidereal rotation on the equator of the Sun
(about 25 days, e.g., see page~77 of~\citealt{Aschwanden} and also~\citealt{Brajsa}; for a survey see~\citealt{rdz}).~For instance depending on the wind speed,
 on the ecliptic the SW spends the corresponding amount of time needed to complete from 12 up to 20 sidereal solar rotations to reach the outer boundary.~In the present code,
the effective heliosphere (with a radius of 100\,AU) was subdivided in 15 spherical regions.~In each region,
 the parameters (e.g., SW speed, $K_0$, $B_\oplus$, $\alpha_{\rm t}$, etc.) are determined at the time of the solar wind ejection.
\section{The Monte Carlo Code HelMod}
\label{Code}
It is worthwhile to note that Eqs.~(\ref{eq_parker},~\ref{eq_parker1}) can be analytically solved only
treating a simplified transport of GCRs through the heliosphere (e.g., see Sect.~\ref{FFModel}
and also~\citealt{gleeson1968,Caballero2004}).~Complex configurations regarding the transport inside the heliosphere
were proposed using numerical methods, like~finite-difference integration~\citep[e.g.,][]{burger1989}.~
\par
As implemented in the HelMod
code\footnote{In the 2D-HelMod code version 1.0, the standard Parker field without drifts was implemented; in version
1.2, the dependence on the particle drift was added; finally, in version
1.4, the Parker magnetic field was modified in polar zones.} version 1.5, the current approach i) follows that from~\cite{yamada1998,gervasi1999,zhang1999,alanko2003,Pei2010,strauss2011}
and ii) exploits a Monte Carlo technique to determine the number density $U$ (Sect.~\ref{Model})
using the set of the approximated stochastic differential equations (SDEs) treated in Appendix~\ref{app1} for a 2-D
approximation (radial distance and colatitude).~For a) an IMF
described by the standard Parker field [Eq.~(\ref{FUN_ParkerField})]  and b) both solar wind and drift
velocity in the region of WHCS radially directed (e.g., $V_{{\rm sw},r}=V_{\rm sw}$ and $v_{{\rm HCS},r}=v_{\rm HCS}$),
the SDEs approximated in terms of the increments $\Delta r$, $\Delta \mu(\theta)$, $\Delta T$ and $\Delta t$
[with $\mu(\theta)\equiv\cos(\theta)$] are (see Appendix~\ref{app1}):
\begin{eqnarray}
% \nonumber to remove numbering (before each equation)
\label{SDEeqs_comp} \Delta r  &=&  \frac{1}{r^2}  \left[\frac{\partial}{\partial r}\left( r^2 K^S_{rr}\right)\right] \Delta t + \left( V_{\rm sw}+ v_{{\rm dr},r}  + v_{\rm HCS} \right)\Delta t + \omega_r\sqrt{2K^S_{rr}\, \Delta t\,} , \\
\label{SDEeqs_comp1}\Delta \mu(\theta) &=& \frac{1}{r^2} \left\{\frac{\partial}{\partial \mu(\theta)}\left\{ [1-\mu^2(\theta)]K^S_{\theta\theta}\right\}\right\} \Delta t - \frac{v_{{\rm dr},\theta} \,\sqrt{1-\mu^2(\theta)}}{r}\,  \Delta t \nonumber\\
                                       & & + \omega_{\mu(\theta)}\sqrt{ \frac{2\,K^S_{\theta \theta}\, [1-\mu^2(\theta)]}{r^2} \Delta t\,}, \\
\label{SDEeqs_comp2}\Delta T  &=&  - \frac{2}{3} \frac{\alpha_{\rm rel} V_{\rm sw} T}{r} \Delta t
\end{eqnarray}
[see Equations~(2--4) of~\citealt{ASTRA_ECRS}, see also~\citealt{Pei2010} and references therein].~For the IMF treated
in Sect.~\ref{HMF} [Eqs.~(\ref{lat_comp}--\ref{B_Kota})] the SDEs~[Eqs.~(\ref{eq::app1}--\ref{eq::app3})] can
be approximated with [Eqs.~(\ref{eq::app7}--\ref{eq::app9})]:
\begin{eqnarray}
\Delta r  &=&\!\left\{\frac{1}{r^2}\frac{\partial}{\partial r}(r^2 K^S_{rr}) - \frac{\partial}{\partial \mu(\theta)}\left[\frac{ K^S_{r\theta}\sqrt{1-\mu^2(\theta)}}{r}\right] + V_{\rm{sw}}+v_{{\rm dr},r}+v_{\rm HCS}\!\right\}\!\Delta t \nonumber\\
           &&    + \omega_r\sqrt{\frac{K^S_{rr}K^S_{\theta\theta}-(K^S_{r\theta})^2}{0.5 K^S_{\theta\theta}}\Delta t\,}   -  \omega_{\mu(\theta)} K^S_{r\theta}\sqrt{\frac{2}{K^S_{\theta\theta}}\,\Delta t\,},
           \label{2D_SDE}   \\
\Delta \mu(\theta) &=&\!\left\{\!-\!\frac{1}{r^2} \frac{\partial}{\partial r}\left(r K^S_{\theta r}\sqrt{1\!-\!\mu^2(\theta)}\right)\!+\! \frac{\partial}{\partial \mu(\theta)}\!\left\{\! \frac{K^S_{\theta \theta}[1\!-\!\mu^2(\theta)]}{r^2}\!\right\}\!-\!\frac{v_{{\rm dr},\theta}\sqrt{1\!-\!\mu^2(\theta)}}{r}\!\right\}\!\Delta t \nonumber\\
           &&  +\omega_{\mu(\theta)}  \sqrt{\frac{2 K^S_{\theta \theta}[1-\mu^2(\theta)]}{r^2} \Delta t\,}, \label{2D_SDE_2}\\
\Delta T  &=&  - \frac{2}{3} \frac{\alpha_{\rm rel} V_{\rm sw} T}{r}\Delta t .\label{2D_SDE_3}
\end{eqnarray}
\par
As discussed by~\citet*{Pei2010} [see also~\citep{strauss2011}], the vector $\vec{q}=(r,\mu,T)$
represents a so-called \textit{pseudoparticle} (see Appendix~\ref{app1}).~Equations~(\ref{2D_SDE}--\ref{2D_SDE_3}) allow
one to simulate the time evolution of pseudoparticles from the outer boundary down to the inner heliosphere.~As
treated by Achterberg and Krulls~(1992), the number density $U$ - or equivalently the differential intensity $J$
[Eq.~(\ref{eq::J})] - can be obtained from \textit{the density of pseudoparticles} by averaging over many realizations of the SDEs.
\par
The procedure to integrate the SDEs is the following:
1) events are isotropically generated  on the outer border of the effective heliosphere;
2) each event is integrated over the time evolution of a pseudoparticle and is processed forward-in-time until
 it reaches either the outer (inner) border of the effective heliosphere located at $100\,$AU ($r_b$) or the pseudoparticle energy
becomes lower than a minimum threshold (which depends on the set of experimental data taken into consideration),
then a new particle is generated;
3) when a pseudoparticle reaches a particular region (for instance that corresponding to Earth position) its
 injection energy, statistical-weight, etc. are recorded;
4) finally, the number density $U$ results from the normalized distribution function %- described in of~\citep{Pei2010} -
obtained using a procedure from~\citet*{Pei2010} (see Section 4.3 in this article).~The forward-in-time approach allows one to reproduce
rigorously processes occurring inside the heliosphere.~
\par
In the present code, $\Delta t$ varies as $r^2 /K_{rr}$, thus allowing an increase of the accuracy in the inner heliosphere,
but keeping the appropriate precision up to regions close to the outer border of the effective heliosphere.~Furthermore,
this condition ensures that the diffusion  process is dominant~\citep[see Section~4.1 of][]{kruell1994}.
\section{Results}
\label{Results}
The current modulation code (Sect.~\ref{Code}) provides a modulated differential intensity for protons using a local interstellar spectrum (LIS) of protons.~In the following, we will discuss i) the LIS used (Sect.~\ref{LIS_discussion}), ii) the comparison of simulated (modulated) differential intensities with those obtained from the measurements of BESS, AMS and PAMELA spectrometers during the solar cycle 23 (Sects.~\ref{H_A_Results},~\ref{R_L_Tilt_Angle}) and iii) the dependence of present results on the treatment of the heliosphere extension (Sect.~\ref{Effective_Hel}).~Furthermore (Sect.~\ref{Heliosph_latid_dep}), the simulated fluxes obtained with HelMod code are compared with (and found to reproduce the features of) the experimental data from Ulysses fast scan in 1995~\citep{simpson1996b}.
\subsection{Local Interstellar Spectrum}
\label{LIS_discussion}
Recently,~\citet*{Herbst2010} reviewed different proton LIS's published in the literature and determined that - as it can be seen in Figure~2(b) in that article - these spectra agree well with each other for proton energies above 10\,GeV%,~i.e., at energies at which the differential intensities of GCRs are expected to be %marginally affected by modulation effects
.~For this comparison, they used, among others, the LIS from Burger, Potgieter and Heber~(2000) (BPH-LIS) in the form of the approximated analytical expression from Usoskin and collaborators~(2005).~Over the past years,
Moskalenko, Strong and collaborators using GALPROP provided a LIS for protons~[e.g.,~see \citet{Galprop3,Galprop2,Galprop1}, see also~\citet{langner2004,2003JGRA_108_8039L}]: the latest calculation agrees with the BPH-LIS above 1\,GV~[e.g.,~see \citet{Galprop1}].~It has to be remarked that the GALPROP spectrum is constrained by a few measured quantities (for instance, the B/C and other isotopes and/or nuclei ratios), some of them will be (accurately) re-determined in the coming years using data from PAMELA and AMS-02 missions.
%%%%%%%%%%%%%%%%%%%%%%%%%%%%%%%%%%%%%%%%%%%%%%%%%%%%
\begin{figure}[t]
\centering
\includegraphics[width=0.48\textwidth]{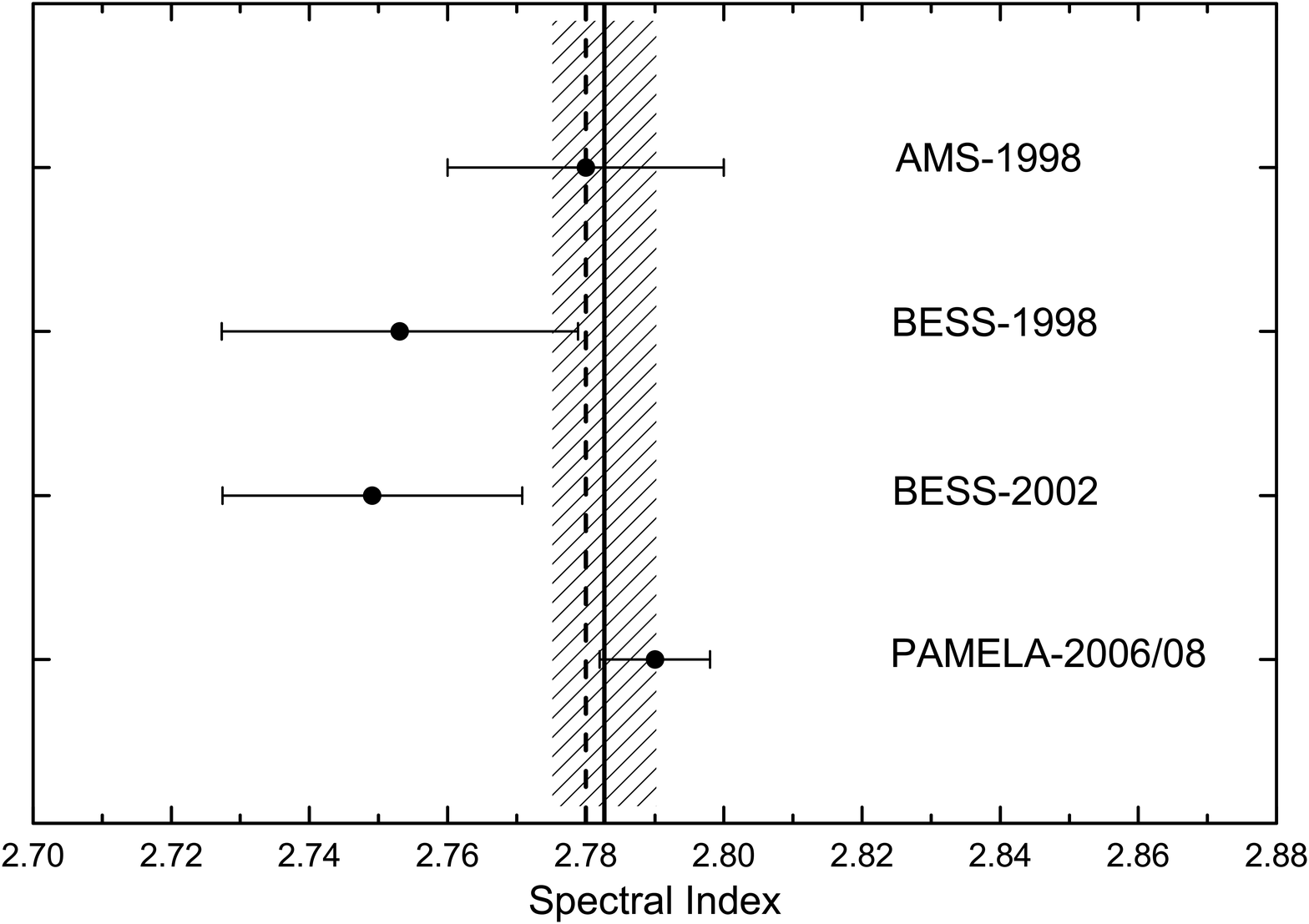}
\includegraphics[width=0.48\textwidth]{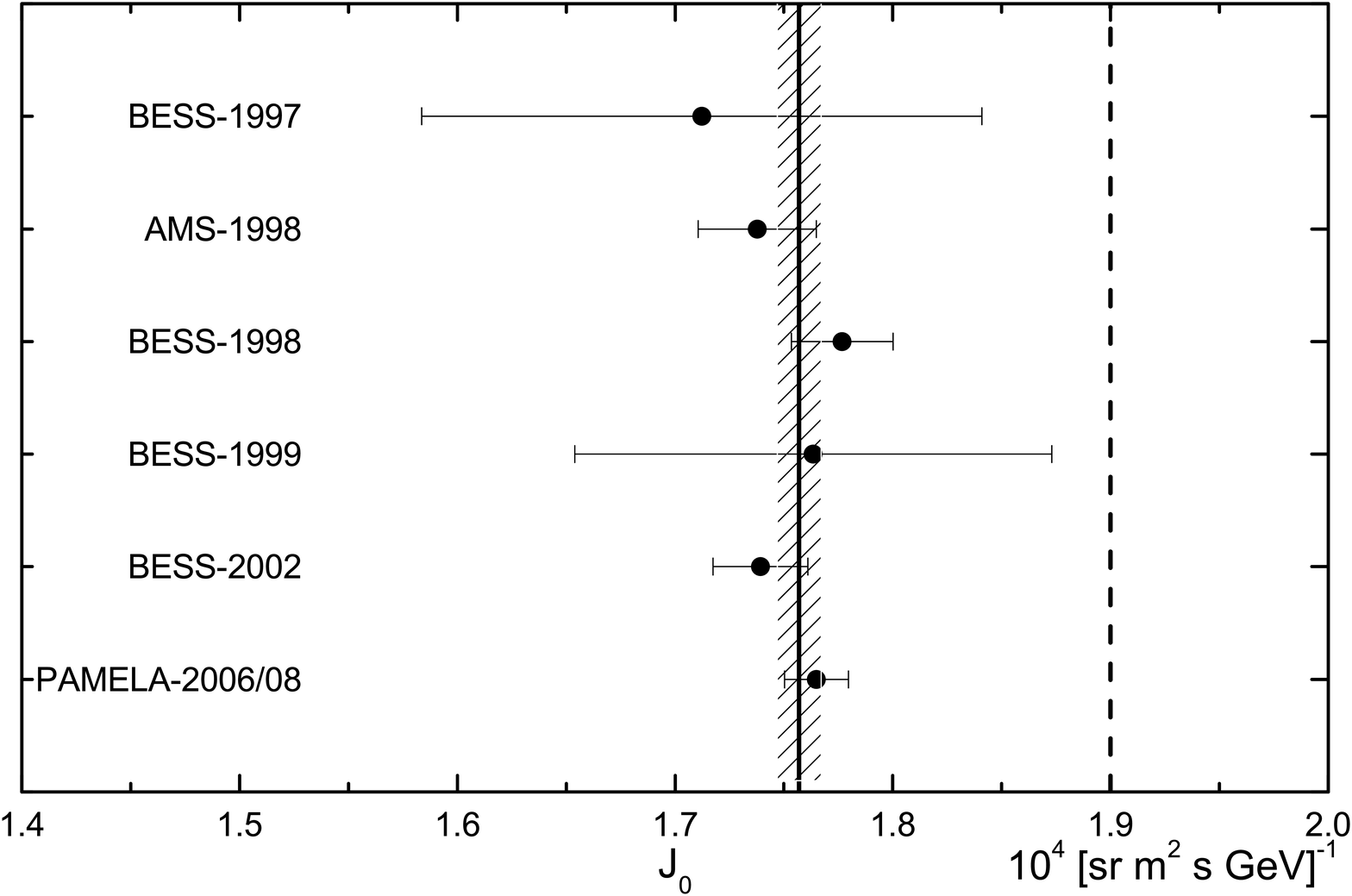}
\caption{Left: spectral index ($\gamma$)
obtained (see text) for %Caprice--1994 \citep{Caprice94},
AMS--1998, BESS--1998, BESS--2002 and
PAMELA--2006/08.~Right: normalization constant ($J_0$) for %Caprice--1994 \citep{Caprice94},
BESS--1997, AMS--1998, BESS--1998, BESS--1999,
%BESS--2000 \citep{BESS_Astropart},
BESS--2002 and PAMELA--2006/08.~The dotted lines represent the values of
the spectral index ($\gamma_{\rm BPH}$) and normalization constant ($J_{0,\rm BHP}$) of the
BPH-LIS, respectively;
the continuous lines represent the error-weighted averages of spectral index ($\gamma_{\rm wa}$) and
normalization constant ($J_{0,\rm wa}$).} \label{fig_simlis}
\end{figure}
%%%%%%%%%%%%%%%%%%%%%%%%%%%%%%%%%%%%%%%%%%%%%%%%%%%%%%%%%%%%%%%%%%
\par
In units of
$\left(\textrm{sr}\,\textrm{m}^2\,\textrm{s}\,\textrm{GeV}\right)^{-1}$ (\citealt{burger2000}, see also~\citealt{ilya02}) the BPH-LIS is expressed as:
\begin{eqnarray}\label{BHH_LIS}
J_{\rm BHP}(T) = \left\{\!\!
\begin{array}{lcc} J_{0,\rm BHP}\, R^{-\gamma_{\rm BPH}}, %& %\qquad
%\qquad\qquad &
\textrm{ for } R  \geq 7 %\qquad
,\\
\\ \exp\!\left[9.472 - 1.999 \,\ln R - 0.6938\,(\ln R)^2  \right. \\
~~~~\left. + 0.2988\,(\ln R)^3 - 0.04714 \,(\ln R)^4 \right], %& \qquad
\qquad
\textrm{ for } %\qquad&
R  < 7 ,
\end{array}\right.
\end{eqnarray}
with
\[
R \equiv R(T) =\frac{ P(T)}{P_0},
\]
where
\[
 P(T) = \frac{\sqrt{T \left( T + 2 E_{\rm r,p} \right)}}{e }
\]
\citep[e.g.,~see Equation~(4.94) in][]{RancBook} is the proton rigidity in GV with $E_{\rm r,p} = m_{\rm p} c^2$, $m_{\rm p}$ is the rest mass of protons in GeV/c$^2$, $T$ is the kinetic energy of proton in GeV, $e$ is
the electron charge, $c$ is the speed of light, $\gamma_{\rm BPH} = 2.78$ is the spectral index, $J_{0,\rm BHP} = 1.9 \times 10^{4}\,\left(\textrm{sr}\,\textrm{m}^2\,\textrm{s}\,\textrm{GeV}\right)^{-1}$ is a normalization constant and, finally, $P_0 = 1\,\textrm{GV}$.
\par
Above (10--20)\,GeV the differential proton intensities are slightly or marginally affected by modulation.~The BPH-LIS [first line of Eq.~(\ref{BHH_LIS})] was compared to experimental spectra available in the literature and collected during the solar cycle 23.~These observations also account for
data in the energy range where modulation is relevant,% and%, at the same time,% within the energy range (20--200)\,GeV,
~e.g.,
AMS--1998 \citep{AMS_tot}, BESS--1998 [with data only in the range (20--117)\,GeV] \citep{sanuki2000},
BESS--2002 \citep{BESS_PLB2} and PAMELA--2006/08 \citep{Pamela_2011}.~In Fig.~\ref{fig_simlis}, the spectral indexes ($\gamma$) of
%Caprice--1994,
AMS--1998 and PAMELA--2006/08 are those from \citep{AMS_tot,Pamela_2011b}, respectively;
while for BESS--1998 and BESS--2002 the spectral indexes were obtained from a fit to the published data of the differential
proton intensities.~It has to be noted that the rigidity independent part of the spectral index found by PAMELA--2006/08 is
%\[
$\gamma_{\rm PAMELA} = 2.790\pm 0.008 \textrm{(stat)}\pm0.001 \textrm{(syst)}$;
%\]
Adriani and collaborators (2011b) found that the spectral index depends on rigidity as expressed
in Equation~(19) therein with a maximum variation of the order of the previously quoted uncertainties in the rigidity range (30--200)\,GV.~Furthermore, the spectral index ($ 2.79 \pm 0.08$) found by Caprice--1994 \citep{Caprice94} is in agreement with those found by the experiments discussed in this section, but the quoted errors are larger.
\par
The normalization constants $J_0$ (Fig.~\ref{fig_simlis}) a) depend on the set of experimental
observations,~e.g., %Caprice--1994 \citep{Caprice94},
BESS--1997 \citep{BESS_Astropart},
BESS--1998 \citep{BESS_Astropart,sanuki2000}, AMS--1998 \citep{AMS_tot}, BESS--1999 \citep{BESS_Astropart}, %BESS--2000 \citep{BESS_Astropart},
BESS--2002 \citep{BESS_PLB2} and PAMELA--2006/08 \citep{Pamela_2011}
and b) were obtained from a fit using $\gamma_{\rm BPH}$ as spectral index to the experimental data.~For BESS--2000 \citep{BESS_Astropart}, the experimental observations did not exceed the 21.5\,GeV, i.e., an energy region of proton differential intensity which might (marginally) still be affected by modulation in a period of high solar activity; thus, the normalization constant used for these data was the one obtained from BESS--2002 \citep{BESS_PLB2} data.
\par
The weighted averages of both the spectral index ($\gamma$) and normalization constant ($J_0$) and their errors were determined following the procedure indicated at pages 14--15 of~\citet{PDB2010}.~The
error-weighted averages found are
\begin{equation}\label{Gammawa}
\gamma_{\rm wa} = 2.783 \pm 0.009
\end{equation}
and
\begin{equation}\label{J0wa}
    J_{0,\rm wa} =(1.76  \pm 0.01)\times 10^4\,\left(\textrm{sr}\,\textrm{m}^2\,\textrm{s}\,\textrm{GeV}\right)^{-1}.
\end{equation}
$\gamma_{\rm wa}$ is well in agreement with that ($\gamma_{\rm BPH}$)
suggested by Burger, Potgieter and Heber~(2000) [Eq.~(\ref{BHH_LIS})].~$J_{0,\rm wa}$ and $\gamma_{\rm wa}$ are represented with the continuous lines in Fig.~\ref{fig_simlis}; in the same figure the dotted lines refer to the values of the BPH-LIS [Eq.~(\ref{BHH_LIS})].~It has to be remarked that the value of $J_0$ found from a fit to Caprice--1994 data above 20\,GeV \citep{Caprice94} is $1.44\pm 0.02$: this value differs by more than 5 standard deviations from $J_{0,\rm wa}$ [Eq.~(\ref{J0wa})].%However, the differences among normalization constants ($J_0$) indicate systematic uncertainties not fully accounted in those quoted.
\par
In Sects.~\ref{H_A_Results}--\ref{R_L_Tilt_Angle}, using the current modulation code the observed proton spectra %-~e.g., the ones from the experiments listed in Fig.~\ref{fig_simlis} and %discussed in this section -
are compared with the modulated differential intensities obtained from
an interstellar differential (per unit of kinetic energy) proton intensity [$J_{\rm HelMod}(T)$] given by
\begin{equation}\label{J_HelMod}
    J_{\rm HelMod}(T) = J_{\rm BHP}(T) \left(\! \frac{J_0}{J_{0,\rm BHP}}\!\right) ~\left[\textrm{sr}\,\textrm{m}^2\,\textrm{s}\,\textrm{GeV}\right]^{-1}.
\end{equation}
$J_{\rm HelMod}(T)$ keeps the same spectral index for $P(T)\geq 7\,$GV as in Eq.~(\ref{BHH_LIS}) and linearly depends on $J_0$, which accounts for the slight absolute fluxes variation among observations.
%%%%%%%%%%%%%%%%%%%%%%%%%%%%%%%%%%%%%%%%%%%%%%%%%%%%%%
\begin{table}[t]
\begin{center}
\begin{tabular}[t]{ccccccc}
&     & ``L'' model & ``R'' model &  \textit{no drift}  & \textit{diagonal approx.} & \textit{scalar approx.}\\
\hline
\hline
BESS--1999  &	&  8.7  &  8.0 &  14.6  &32.0 &   29.7 \\
BESS--2000 &     &   16.2  &  15.8 &  13.0 & 23.6 &26.7 \\
BESS--2002 &     &  12.7  &  15.0 & 12.2 & 34.8 & 33.2 \\
\hline
\end{tabular}
\caption{For BESS--1999, BESS--2000 and BESS--2002, $\eta_{\rm RMS}$ (in percentage) obtained from Eq.~(\ref{eta_rms})
\textit{with enhancement of the diffusion tensor along the polar direction}
using ``L'' and ``R'' models for the tilt angle %($\alpha_{\rm t}$) [Sect.~\ref{dynamic} and~\citep{Hoek95,wsoWeb}]
and for \textit{no drift} approximation, \textit{diagonal approximation} and, finally, \textit{scalar approximation} (see text).~%The simulated differential intensity are obtained with an enhancement of the diffusion tensor along %the polar direction ($K_{\perp \theta}$) [Eq.~(\ref{enhaced_K})] and inside a heliospheric region where latitudinal magnitudes correspond to angles (from the ecliptic plane) lower than $|30^\circ|$.
}\label{table::max}
%\end{center}
%\end{table}
%%%%%%
\vskip +0.5cm
%\begin{table}
%\begin{center}
\begin{tabular}[t]{ccccccc}
&     & ``L'' model & ``R'' model &  \textit{no drift}  & \textit{diagonal approx.} & \textit{scalar approx.}\\
\hline
\hline
BESS--1999&	&   6.8 &  8.1 &  24.3  &  30.5  & 30.2  \\
BESS--2000&     &  11.3 & 10.2 & 10.8 &  26.2 &  26.2  \\
BESS--2002&     & 13.0 & 15.7 & 12.7 & 33.9 & 33.2 \\
\hline

\end{tabular}
\caption{For BESS--1999, BESS--2000 and BESS--2002, $\eta_{\rm RMS}$ (in percentage) obtained from Eq.~(\ref{eta_rms}) \textit{without any enhancement of the diffusion tensor along the polar direction}
using ``L'' and ``R'' models for the tilt angle %($\alpha_{\rm t}$) [Sect.~\ref{dynamic} and~\citep{Hoek95,wsoWeb}]
and for \textit{no drift} approximation, \textit{diagonal approximation} and, finally, \textit{scalar approximation} (see text).~%The simulated differential intensity are obtained with $K_{\perp %\theta}=K_{\perp r}$ independently %of the latitude and inside a heliospheric region where latitudinal magnitudes correspond to angles (from the ecliptic plane) lower than $|30^\circ|$.
}\label{table::max_noen}
\end{center}
\end{table}
%%%%%%
%
%
\subsection{Comparison with Observations Obtained During Solar Cycle 23}
\label{H_A_Results}
We used the present code for quantitative comparisons [using Eqs.~(\ref{eta_rms},~\ref{eta_i})] with experimental data (discussed later in this section) collected during solar cycle 23, in periods with \textit{high solar activity},~i.e., when the solar magnetic field becomes increasingly complex and less dipolar (Sects.~\ref{Model},~\ref{HMF}).~This code allowed us to investigate how the modulated (simulated) differential intensities are affected by the i) particle drift effect (Sects.~\ref{Model},~\ref{antysymmetric_part}), ii) polar enhancement of the diffusion tensor along the polar direction ($K_{\perp \theta}$)  [Eq.~(\ref{enhaced_K})] and, finally, iii) the value of tilt angles ($\alpha_{\rm t}$) calculated following the approach due to ``R'' and ``L'' models [Sect.~\ref{dynamic} and~\citep{Hoek95,wsoWeb}].~The
magnetic field is modified with respect to Parker's magnetic field in the polar region as proposed by~\cite{JokipiiKota89} (Sect.~\ref{HMF}).~
%%%%%%%%%%%%%%%%%%%%%%%%%%%%%%%%%%%%%%%%%%%%%%%%%%%%%%%%%%%%%%%%%%%%%%%%%%%%%%%%%%%%%%%
 \begin{figure}[t]
   \vskip-0.5cm
 \begin{center}
  \includegraphics[width=0.5\textwidth]{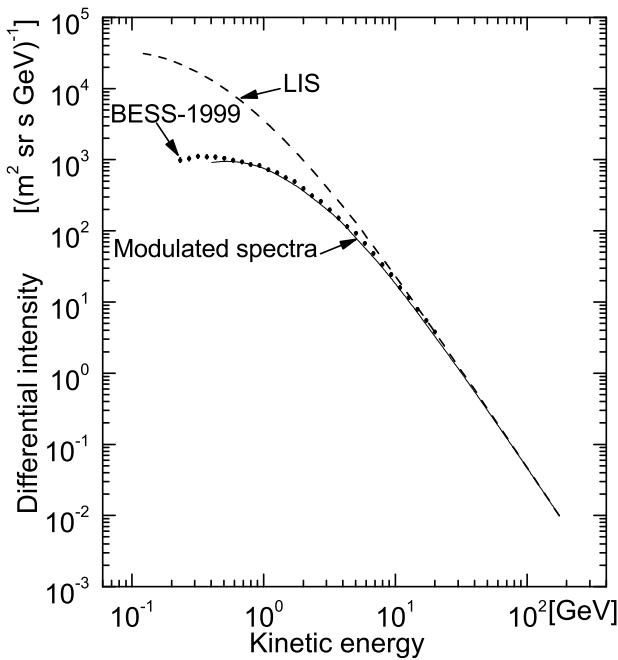}
  \vskip-0.5cm
  \caption{Differential intensity determined with HelMod code (continuous line) %for a heliospheric region where latitudinal magnitudes correspond to angles (from the ecliptic plane) lower than $|30^\circ|$
  compared to the experimental data of BESS--1999%~\citep{BESS_Astropart}
  ; the dashed line is the LIS (see text).~%The modulated differential intensities was calculated using $K_{\perp \theta}=K_{\perp r}$ independently of the %latitude and including the charge %drift effect with the %values of tilt angle from ``L'' model.~The dashed line is the LIS [Eqs.~(\ref{BHH_LIS},~\ref{J_HelMod})] with normalization constant $J_0$ corresponding to BESS-1999 %(Sect.\ref{LIS_discussion}).
  }\label{Fig:Bess_1999}
% \vskip-0.5cm
\end{center}
\end{figure}
% %%%%%%%%%%%%%%%%%%%%%%%%%%%%%%%%%%%%%%%%%%%%%%%%%%%%%%%%%%%%%%%%%%%%%%%%%%%%%%%
\par
The effects related to particle drift were investigated (a) via the suppression of the drift velocity
-~i.e., under the assumption that $K_A=0$ (Sect.~\ref{antysymmetric_part}), thus \textit{no drift} convection was accounted for -, (b) in a pure diffusion approximation with a diagonal diffusion
tensor (termed \textit{diagonal approximation}), where $K_{rr}=\mathcal{K}$ and $K_{\theta\theta}=\rho_k \mathcal{K} $ (Sect.~\ref{Model}) and, finally, (c) in a pure diffusion approximation with components both equal to  $\mathcal{K}$ (called \textit{scalar approximation}) [as in Eq.~(\ref{diffusion_coeff})].~The case (a) accounts the hypothesis that magnetic drift convection is almost completely suppressed during solar maxima.~In addition, for cases (b) and (c) one allows to assume that the diffusion propagation is independent of magnetic structure.
% %%%%%%%%%%%%%%%%%%%%%%%%%%%%%%%%%%%%%%%%%%%%%%%%%%%%%%%%%%%%%%%%%%%%%%%%%%%%%%%%%%%%%%%
\begin{figure}[t]
\begin{center}
  \includegraphics[width=0.5\textwidth]{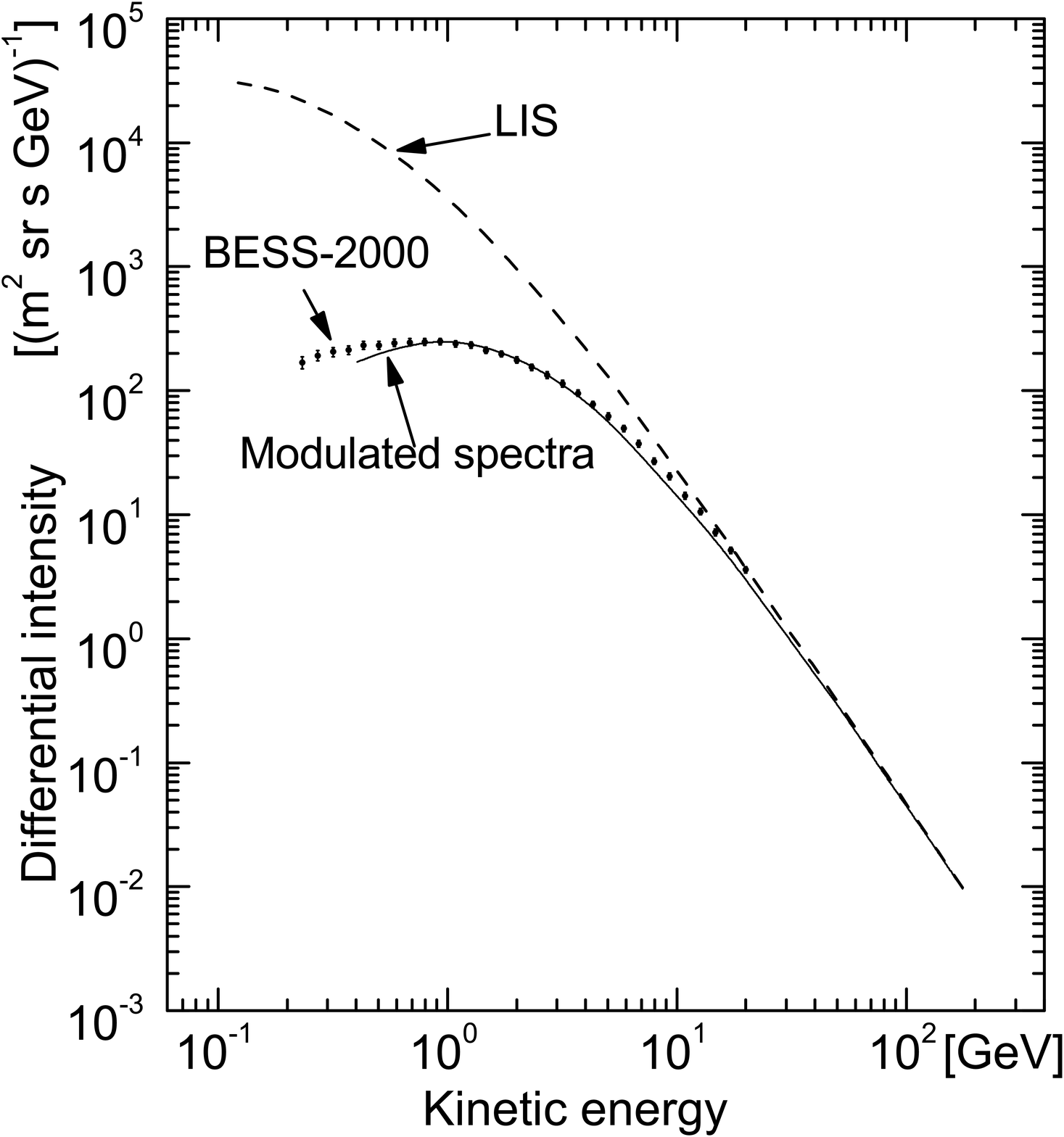}
   \vskip-0.5cm
  \caption{Differential intensity determined with HelMod code (continuous line) %for a heliospheric region where latitudinal magnitudes correspond to angles (from the ecliptic plane) lower than $|30^\circ|$
  compared to the experimental data of BESS--2000%~\citep{BESS_Astropart}
  ; the dashed line is the LIS (see text).~%The modulated differential intensities was calculated using $K_{\perp \theta}=K_{\perp r}$ independently of the %latitude and including the charge %drift %effect with the values of tilt angle from ``L'' model.~The dashed line is the LIS [Eqs.~(\ref{BHH_LIS},~\ref{J_HelMod})] with normalization constant $J_0$ corresponding to BESS-2002 %(Sect.\ref{LIS_discussion}).
  }\label{Fig:Bess_2000}
% \vskip-0.5cm
\end{center}
\end{figure}
% %%%%%%%%%%%%%%%%%%%%%%%%%%%%%%%%%%%%%%%%%%%%%%%%%%%%%%%%%%%%%%%%%%%%%%%%%%%%%%%
% %%%%%%%%%%%%%%%%%%%%%%%%%%%%%%%%%%%%%%%%%%%%%%%%%%%%%%%%%%%%%%%%%%%%%%%%%%%%%%%%%%%%%%%
\begin{figure}[t]
\begin{center}
  \includegraphics[width=0.5\textwidth]{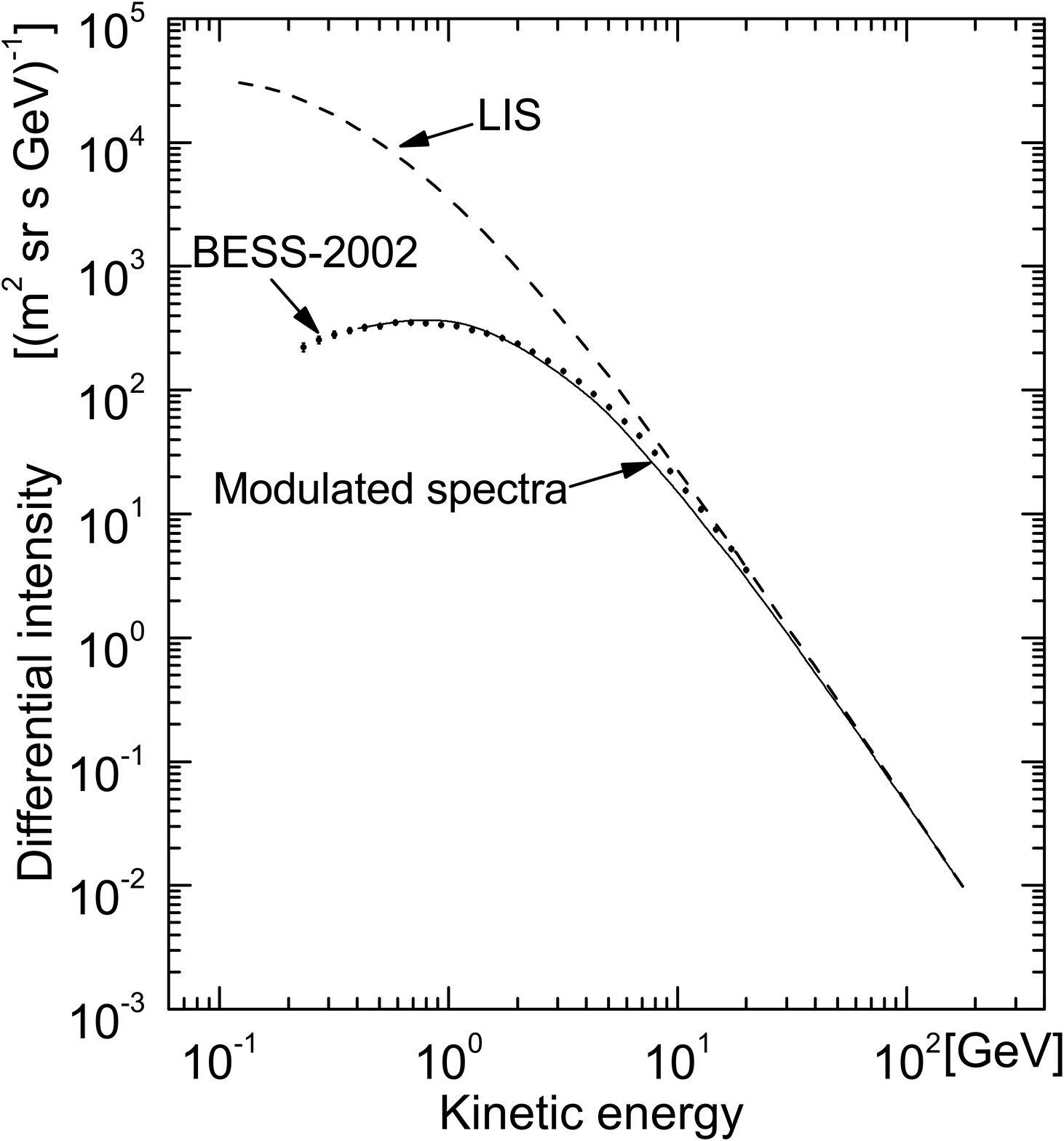}
  \vskip-0.5cm
  \caption{Differential intensity determined with HelMod code (continuous line) %for a heliospheric region where latitudinal magnitudes correspond to angles (from the ecliptic plane) lower than $|30^\circ|$
  compared to the experimental data of BESS--2002%~\citep{BESS_Astropart}
  ; the dashed line is the LIS (see text).%~The modulated differential intensities was calculated using $K_{\perp \theta}=K_{\perp r}$ independently of the %latitude and  including the charge %drift effect with the values of tilt angle from ``L'' model.~The dashed line is the LIS [Eqs.~(\ref{BHH_LIS},~\ref{J_HelMod})] with normalization constant $J_0$ corresponding to BESS-2002 %(Sect.\ref{LIS_discussion}).
  }\label{Fig:Bess_2002}
 \end{center}
 \end{figure}
%%%%%%%%%%%%%%%%%%%%%%%%%%%%%%%%%%%%%%%%%%%%%%%%%%%%%%%%%%%%%%%%%%%%%%%%%%%%%%%
\par
Each modulated (simulated) differential intensity was obtained using a diffusion tensor (Sects.~\ref{Model},~\ref{FFModel},~\ref{antysymmetric_part} and Appendix~\ref{app1}), whose elements depend on the actual value of the diffusion parameter $K_0$.~Furthermore, the modulated spectra were derived from a LIS [Eqs.~(\ref{BHH_LIS},~\ref{J_HelMod})] whose normalization constant ($J_0$) depends on the experimental set of data (see discussion in Sect.~\ref{LIS_discussion}).~In addition, these differential intensities were calculated 1) for a polar-increased value of $K_{\perp \theta}$ [Eq.~(\ref{enhaced_K})] and also with
$K_{\perp \theta}=K_{\perp r}$, and 2) accounting for particles inside two heliospheric regions where solar latitudes are lower than $|5.7^\circ|$ and $|30^\circ|$, respectively.~As discussed in Sect.~\ref{FFModel}, in
the present model $K_P$ is assumed to be equal to the value of the
rigidity ($P$) [Eq.~(\ref{KP_Linear})] above proton kinetic
energies of $\approx 444\,$MeV~\citep[e.g.,][]{gloek66,gleeson1968,perko1987,potgieter94}.~However, it has to be remarked that a systematic investigation of its dependence below that value and the shape of low energy part of the LIS spectrum [Eqs.~(\ref{BHH_LIS},~\ref{J_HelMod})] was not attempted using the modulated intensities obtained from HelMod code.~In fact, this investigation is likely to be carried out using the experimental data from accurate observations over a long duration, like those from the AMS-02 spectrometer which will allow one to reconstruct the particle trajectory.~The reconstructed particle trajectory results in untangling GCRs coming from outside the magnetosphere also at large geomagnetic latitudes ($\Theta_{\rm M}$) where less energetic particles can enter the magnetosphere.~For instance, inside highest geomagnetic region with $ 1 < \Theta_{\rm M} < 1.1$\,radian [e.g.,~see Figure~2(c) in~\citep{AMS_protons} and Figure~8 in~\citep{mi_jgr}] AMS-1998 data indicate that i) the effective geomagnetic cut-off prevents primary protons (i.e.,~CR protons) from being fully observed with energies below $\approx (0.5$--0.6)\,GeV and ii) secondary particles largely contribute to the overall differential intensity.~In addition, it has be noted that BESS observations were usually performed at large geomagnetic latitudes with $\Theta_{\rm M}$ close to 1.13\,radian.
\par
The past period of high solar activity was during the solar cycle 23; BESS collaboration took data in the years 1999, 2000 and 2002 [see sets of data in~\citep{BESS_Astropart}].~These data were compared with those obtained by means of HelMod code using the error-weighted root mean square ($\eta_{\rm RMS}$) of the
relative difference ($\eta$) between experimental data ($f_{\rm exp}$) and those resulting from simulated differential intensities ($f_{\rm sim}$).~For each set of experimental data and above described approximations and/or models, we determined the quantity:
\begin{equation}\label{eta_rms}
 \eta_{\rm RMS}=\sqrt{\frac{\sum_i \left(\eta_i/\sigma_{\eta,i}\right)^2}{\sum_i 1/\sigma^2_{\eta,i}} }
\end{equation}
with
\begin{equation}\label{eta_i}
 \eta_i=\frac{ f_{\rm sim}(T_i) -f_{\rm exp}(T_i) }{f_{\rm exp}(T_i)},
\end{equation}
where $T_i$ is the average energy of the $i$-th energy bin of the differential intensity distribution and $\sigma_{\eta,i}$ are the errors including the experimental and Monte Carlo uncertainties; the latter account
for the Poisson error of each energy bin.~The simulated differential intensities are interpolated with a cubic spline function.~
\par
In Tables~\ref{table::max},~\ref{table::max_noen}, the values of the parameter $\eta_{\rm RMS}$ (in percentage) are shown; they were obtained in the energy range\footnote{Above 30\,GeV the differential intensity is marginally (if at all) affected by modulation.} from 444\,MeV up to 30\,GeV using ``L'' and ``R'' models for the tilt angle ($\alpha_{\rm t}$) [Sect.~\ref{dynamic} and~\citep{Hoek95,wsoWeb}], for \textit{no drift} approximation, \textit{diagonal approximation} and \textit{scalar approximation} (approximations discussed previously in this section), finally with (Table~\ref{table::max}) and without (Table~\ref{table::max_noen}) the enhancement of the diffusion tensor along the polar direction ($K_{\perp \theta}$) [Eq.~(\ref{enhaced_K})].~The simulated differential intensity were obtained for a heliospheric region where solar latitudes are lower than $|30^\circ|$.~From inspection of Tables~\ref{table::max} and~\ref{table::max_noen}, one can note that i) the \textit{no drift approximation} is better appropriate than \textit{diagonal} and \textit{scalar approximations}, ii) %(when the drift velocity is taken into account)
the ``L''  model for calculating the values of tilt angle ($\alpha_{\rm t}$) is slightly to be preferred to ``R'' model (although the overall differences between these two models are marginal), iii) the results obtained accounting for drift effects using tilt angles from ``L'' model are better in agreement with experimental data with respect to the \textit{no drift} approximation and, finally, iv) the minimum difference with the experimental data occurs when $K_{\perp \theta}=K_{\perp r}$ is assumed independently of the latitude (Table~\ref{table::max_noen}, see first column of the left-hand side).~In addition, the results obtained for a heliospheric region where solar latitudes are lower than $|5.7^\circ|$ exhibit a behavior similar to those lower than $|30^\circ|$, but with values of $\eta_{\rm RMS}$ (in percentage) larger by about several percents.~In Figs.~\ref{Fig:Bess_1999},~\ref{Fig:Bess_2000},~\ref{Fig:Bess_2002} the differential intensities determined with HelMod code are shown and compared with the experimental data of BESS--1999, BESS--2000 and BESS--2002, respectively; in the same figures, the dashed line is the LIS [Eqs.~(\ref{BHH_LIS},~\ref{J_HelMod})] with normalization constants $J_0$ treated in Sect.\ref{LIS_discussion}.~These modulated intensities are the ones calculated for a heliospheric region where solar latitudes are lower than $|30^\circ|$, using $K_{\perp \theta}=K_{\perp r}$ independently of the latitude and including particle drift effects with the values of tilt angle from the ``L'' model.
\par
Finally, it has be concluded that the present code combining diffusion and drift mechanisms is suited to describe the modulation effect in periods with high solar activity \citep[e.g.,~see][]{FER_POT,Ndtiitwani}.
\subsubsection{Periods not Dominated by High Solar Activity}
\label{R_L_Tilt_Angle}
%
%%%%%%%%%%%%%%%%%%%%%%%%%%%%%%%%%%%%%%%%%%%%%%%%%%%%%%%%%%%%%%%%%%%%%%%%%%%%%%%%%%%%%%%%%%%%%%%%%%%%%%%%%%%%%%%%%%%%%%%%%%%%%%%%%%%%%%%%%%%%%%%%%%%%%%%%%%%%%%%%%%%%%%%%%
\begin{table}[t]
%\vspace{ -1cm}
\begin{center}

%\resizebox{0.5\textwidth}{!}{
 \begin{tabular}[t]{ccccccc}
    &     & ``L'' model & ``R'' model &  \textit{no drift}  & \textit{diagonal approx.} & \textit{scalar approx.}\\
\hline
\hline
BESS--1997    &     &  9.2 & 17.7 &  10.4 & 9.5 &  17.6  \\
\hline
AMS--1998 &	&  4.6 & 7.9 &  12.9  & 5.4 &  17.3 \\
\hline
BESS--1998 &	& 9.1 & 14.1 & 9.3   &  4.7 &  13.6 \\
\hline
PAMELA--2006/08 &     &  7.1 &  13.4 & 5.9  & 17.5 & 52.5 \\
\hline
\end{tabular}
\caption{For BESS--1997, AMS--1998, BESS--1998 and PAMELA--2006/08, $\eta_{\rm RMS}$ (in percentage) obtained from Eq.~(\ref{eta_rms}) \textit{with enhancement of the diffusion tensor along the polar direction}
using ``L'' and ``R'' models for the tilt angle %($\alpha_{\rm t}$) [Sect.~\ref{dynamic} and~\citep{Hoek95,wsoWeb}]
and for \textit{no drift} approximation, \textit{diagonal approximation} and, finally, \textit{scalar approximation} (see Sect.~\ref{H_A_Results}).~%The simulated differential intensity are obtained with an enhancement of the %diffusion tensor along the polar direction ($K_{\perp \theta}$) [Eq.~(\ref{enhaced_K})] and inside a heliospheric region where latitudinal magnitudes correspond to angles (from the ecliptic plane) lower than $|5.7^\circ|$.
}\label{table::min}
%\end{center}
%\end{table}
\vskip +0.5cm
%\begin{table}[htb!]
%\begin{center}
\begin{tabular}[t]{ccccccc}
    &     & ``L'' model & ``R'' model &  \textit{no drift}  & \textit{diagonal approx.} & \textit{scalar approx.}\\
\hline
\hline
BESS--1997&     &  13.4 &  20.6 &  14.2 &  11.13 & 12.0 \\
\hline
AMS--1998 &     &  6.1 &  11.3 &  11.4 &  6.0 &  3.7 \\
\hline
BESS--1998&     &  11.1 & 17.7 &  7.3 &  4.1 & 7.1 \\
\hline
PAMELA--2006/08 &     &  11.0 &  24.7  &  5.4 &  12.3 &  30.6 \\
\hline
\end{tabular}
\caption{For BESS--1997, AMS--1998, BESS--1998 and PAMELA--2006/08, $\eta_{\rm RMS}$ (in percentage) obtained from Eq.~(\ref{eta_rms}) \textit{without any enhancement of the diffusion tensor along the polar direction} using  ``L'' and ``R'' models for the tilt angle %($\alpha_{\rm t}$) [Sect.~\ref{dynamic} and~\citep{Hoek95,wsoWeb}]
and for \textit{no drift} approximation, \textit{diagonal approximation} and, finally, \textit{scalar approximation} (see Sect.~\ref{H_A_Results}).~%The simulated differential intensity are obtained with $K_{\perp %\theta}=K_{\perp %r}$ independently of the latitude and inside a heliospheric region where latitudinal magnitudes correspond to angles (from the ecliptic plane) lower than $|5.7^\circ|$.
}\label{table::min_noen}
\end{center}
\end{table}
%%%%%%%%%%%%%%%%%%%%%%%%
%%%%%%%%%%%%%%%%%%%%%%%%%%%%%%%%%%%%%%%%%%%%%%%%%%%%%%%
 \begin{figure}[t]
 \begin{center}
  \includegraphics[width=0.5\textwidth]{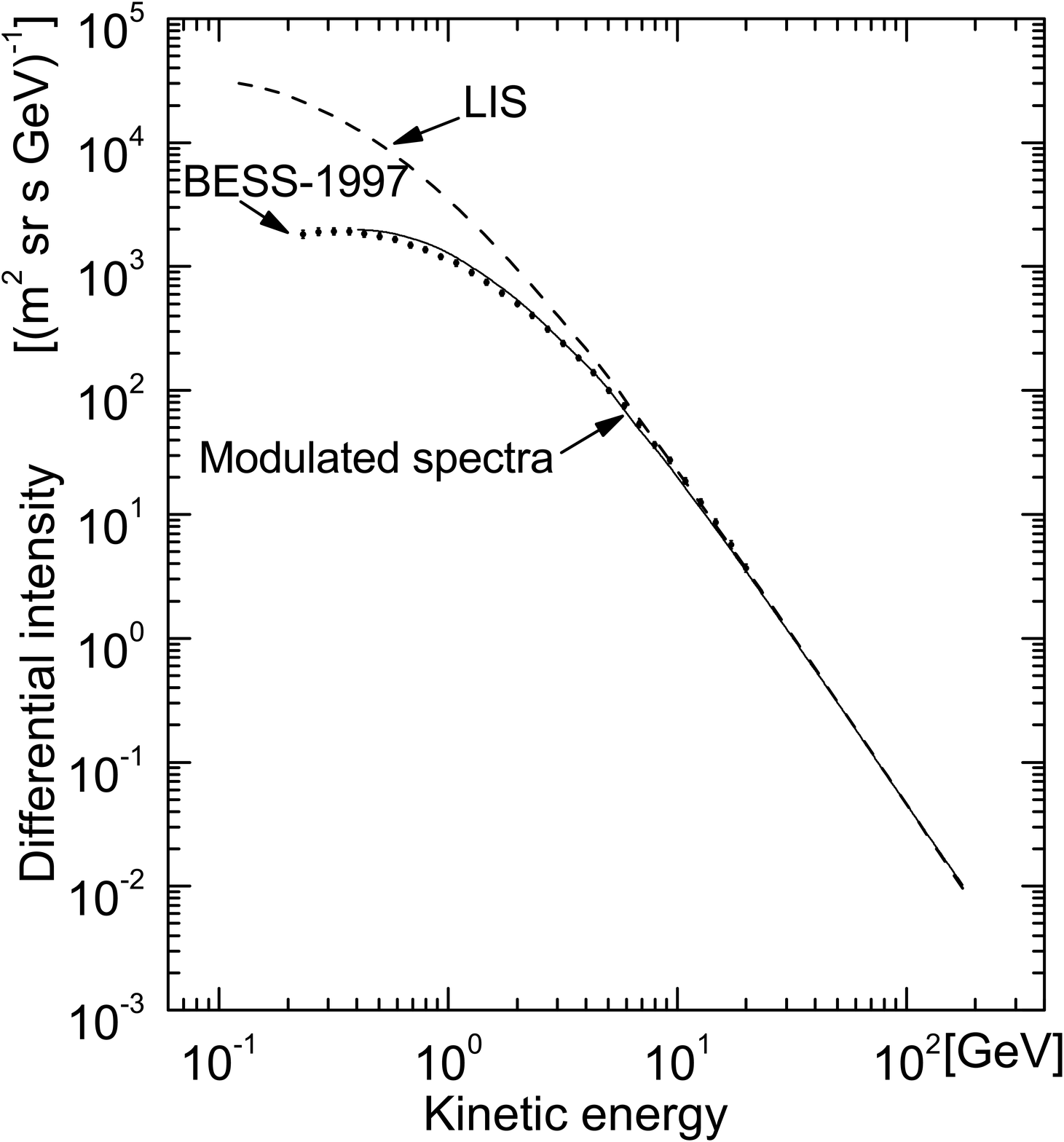}
  \vskip-0.5cm
 \caption{Differential intensity determined with HelMod code (continuous line) %for a heliospheric region where latitudinal magnitudes correspond to angles (from the ecliptic plane) lower than $|5.7^\circ|$
 compared to the experimental data of BESS--1997%~\citep{BESS_Astropart}
 ; the dashed line is the LIS (see text).~%The modulated differential intensities was calculated using the enhancement of the diffusion tensor along the polar  %direction ($K_{\perp \theta}$) % [Eq.~(\ref{enhaced_K})] and including the charge drift effect with the values of tilt angle from ``L'' model.~The dashed line is the LIS [Eqs.~(\ref{BHH_LIS},~\ref{J_HelMod})] with  %normalization  %constant $J_0$ corresponding to BESS--1997 (Sect.\ref{LIS_discussion}).
 }\label{Fig:Bess_1997}
\end{center}
\end{figure}
%%%%%%%%%%%%%%%%%%%%%%%%%%%%%%%%%%%%%%%%%%%%%%%%%%%
\begin{figure}[t]
\begin{center}
    \includegraphics[width=0.5\textwidth]{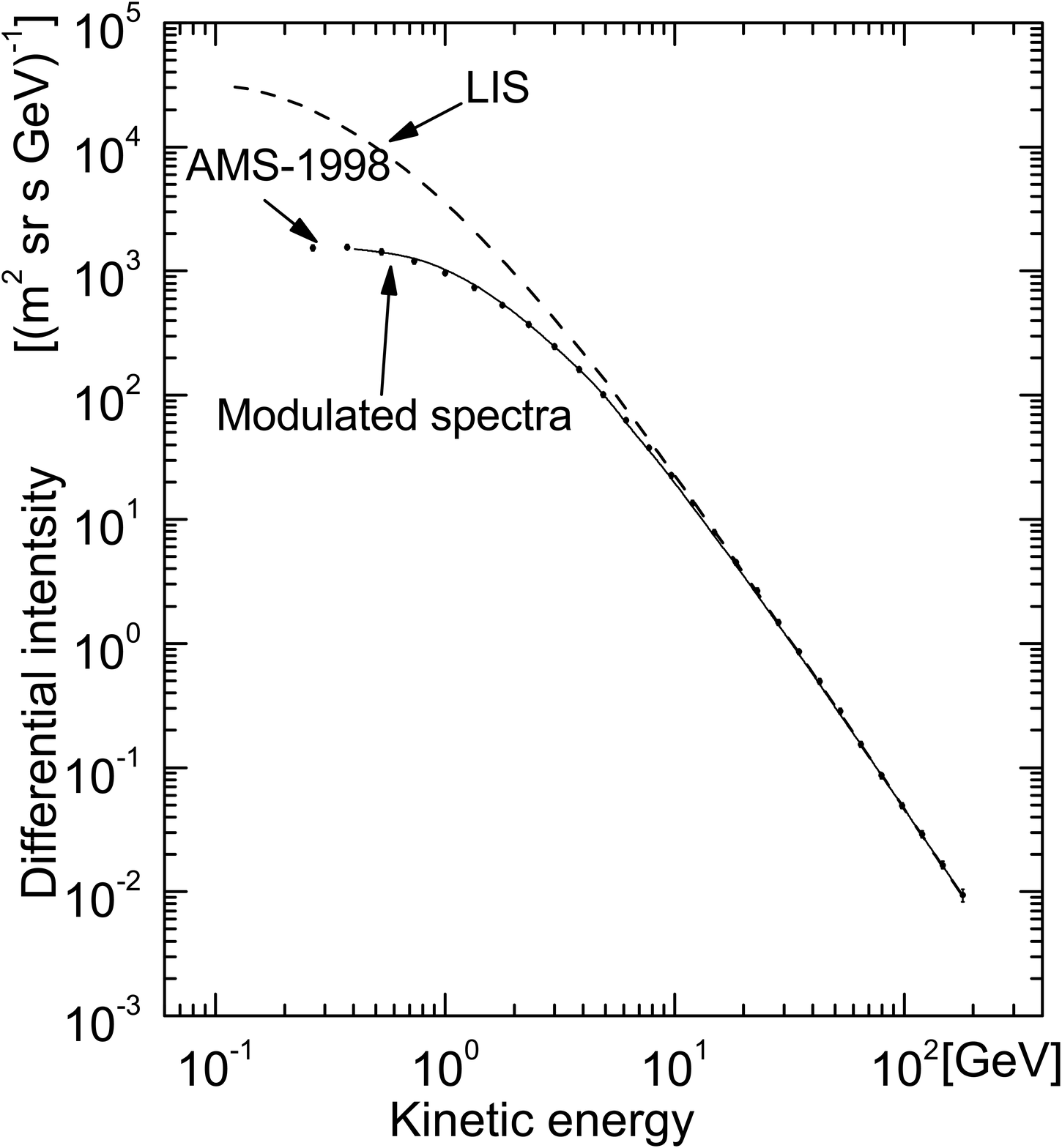}
  \vskip-0.5cm
\caption{Differential intensity determined with HelMod code (continuous line) %for a heliospheric region where latitudinal magnitudes correspond to angles (from the ecliptic plane) lower than $|5.7^\circ|$
compared to the experimental data of AMS--1998%~\citep{AMS_tot}
; the dashed line is the LIS (see text).~%The modulated differential intensities was calculated using the enhancement of the %diffusion tensor along the polar direction %($K_{\perp \theta}$) [Eq.~(\ref{enhaced_K})] and including the charge drift effect with the values of tilt angle from ``L'' model.~The dashed line %is the LIS [Eqs.~(\ref{BHH_LIS},~\ref{J_HelMod})] with normalization constant $J_0$ %corresponding to AMS--1998 (Sect.\ref{LIS_discussion}).
} \label{Fig:ams}
 \end{center}
 \end{figure}
% %%%%%%%%%%%%%%%%%%%%%%%%%%%%%%%%%%%%%%%%%%%%%%%%%%%%%%%%%%%%%%%%%%%%%%%%%%%%%%%%%%%%%%%
%%%%%%%%%%%%%%%%%%%%%%%%%%%%%%%%%%%%%%%%%%%%%
 \begin{figure}[t]
 \begin{center}
\includegraphics[width=0.5\textwidth]{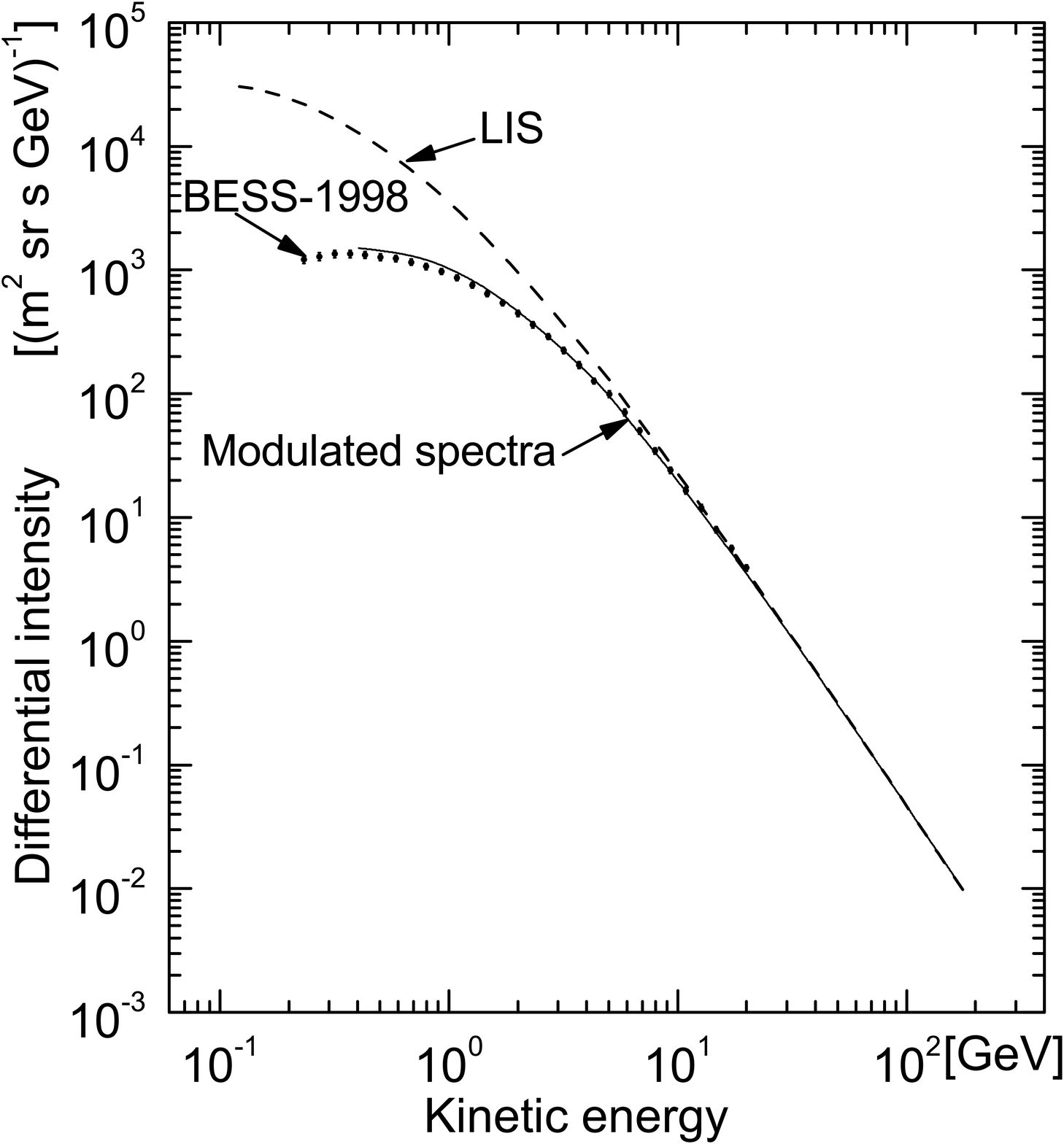}
   \vskip-0.5cm
\caption{Differential intensity determined with HelMod code (continuous line) %for a heliospheric region where latitudinal magnitudes correspond to angles (from the ecliptic plane) lower than $|5.7^\circ|$
compared to the experimental data of BESS--1998%~\citep{BESS_Astropart}
; the dashed line is the LIS (see text).~%The modulated differential intensities was calculated using the enhancement of %the diffusion tensor along the polar %direction ($K_{\perp \theta}$) [Eq.~(\ref{enhaced_K})] and including the charge drift effect with the values of tilt angle from ``L'' model.~The dashed %line is the LIS [Eqs.~(\ref{BHH_LIS},~\ref{J_HelMod})] with normalization %constant $J_0$ corresponding to BESS--1998 (Sect.\ref{LIS_discussion}).
}\label{Fig:Bess_1998}
\end{center}
\end{figure}
% %%%%%%%%%%%%%%%%%%%%%%%%%%%%%%%%%%%%%%%%%%%%%%%%%%%%%%%%%%%%%%%%%%%%%%%%%%%%%%%
% %%%%%%%%%%%%%%%%%%%%%%%%%%%%%%%%%%%%%%%%%%%%%%%%%%%%%%%%%%%%%%%%%%%%%%%%%%%%%%%%%%%%%%%
\begin{figure}[t]
\begin{center}
  \includegraphics[width=0.5\textwidth]{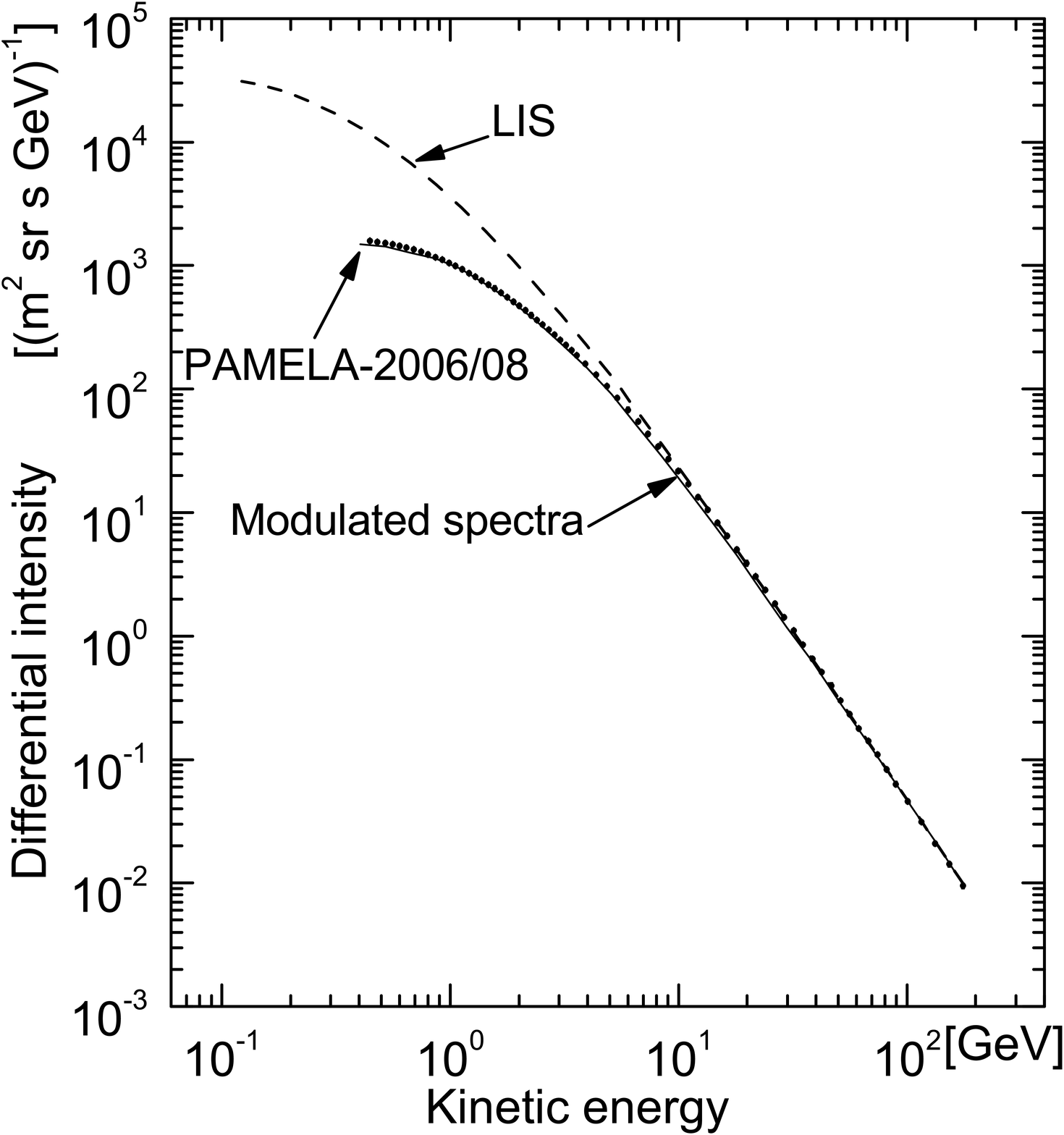}
   \vskip-0.5cm
\caption{Differential intensity determined with HelMod code (continuous line) %for a heliospheric region where latitudinal magnitudes correspond to angles (from the ecliptic plane) lower than $|5.7^\circ|$
compared to the experimental data of PAMELA--2006/08%~\citep{Pamela_2011}
; the dashed line is the LIS (see text).~%The modulated differential intensities was calculated using the enhancement %of the diffusion tensor along the polar %direction ($K_{\perp \theta}$) [Eq.~(\ref{enhaced_K})] and including the charge drift effect with the values of tilt angle from ``L'' model.~The %dashed line is the LIS [Eqs.~(\ref{BHH_LIS},~\ref{J_HelMod})] with normalization %constant $J_0$ corresponding to PAMELA--2006/08 (Sect.\ref{LIS_discussion}).
} \label{Fig:pamela}
 \end{center}
 \end{figure}
%%%%%%%%%%%%%%%%%%%%%%%%%%%%%%%%%%%%%%%%%%%%%%%%%%%%%%%%%%%%%%%%%%%%%%%%%%%%%%%
In periods where the \textit{solar activity is no longer at maximum}, the solar magnetic field becomes increasingly dipolar (Sects.~\ref{Model},~\ref{HMF}).~We used the present code to compare the simulated differential intensities with experimental data obtained during periods not dominated by high solar activity in the solar cycle 23,~i.e., BESS--1997 \citep{BESS_Astropart}, AMS--1998 \citep{AMS_tot}, BESS--1998 \citep{BESS_Astropart,sanuki2000} and PAMELA--2006/08 \citep{Pamela_2011}.~As discussed in Sect.~\ref{H_A_Results}, the simulated spectra were calculated including the effects due to particle drift - expected to be relevant (Sects.~\ref{Model},~\ref{antysymmetric_part}) - with the value of tilt angles ($\alpha_{\rm t}$) calculated following the approach due to ``R'' and ``L'' models [Sect.~\ref{dynamic} and~\citep{Hoek95,wsoWeb}], with and without the polar enhancement of the diffusion tensor along the polar direction ($K_{\perp \theta}$) [Eq.~(\ref{enhaced_K})].~Similarly to the treatment for periods with high solar activity (Sect.~\ref{H_A_Results}), the effects related to particle drift were also investigated (a) via the suppression of the drift velocity (\textit{no drift}), (b) with the \textit{diagonal approximation} and, finally, (c) with the \textit{scalar approximation}.
\par
In Tables~\ref{table::min} and~\ref{table::min_noen}, the values of the parameter $\eta_{\rm RMS}$ (in percentage) are shown.~They were obtained in the energy range from 444\,MeV up to 30\,GeV using ``L'' and ``R'' models for the tilt angle ($\alpha_{\rm t}$) [Sect.~\ref{dynamic} and~\citep{Hoek95,wsoWeb}], for \textit{no drift} approximation, \textit{diagonal approximation} and \textit{scalar approximation} (approximations discussed in this Sect.\ref{H_A_Results}), finally with (Table~\ref{table::min}) and without (Table~\ref{table::min_noen}) the enhancement of the diffusion tensor along the polar direction ($K_{\perp \theta}$) [Eq.~(\ref{enhaced_K})].~The simulated differential intensity were obtained for a heliospheric region where solar latitudes are lower than $|5.7^\circ|$.~From inspection of Tables~\ref{table::min} and~\ref{table::min_noen}, one can note that i) the \textit{diagonal approximation} is better appropriate than \textit{no drift} and \textit{scalar approximations}, ii) %(when the drift velocity is taken into account)
the ``L''  model %for calculating the values of
for tilt angles ($\alpha_{\rm t}$)
is slightly to be preferred to ``R'' model %(similarly to the results discussed in Sect.~\ref{H_A_Results}) %iii) the results obtained using the charge effect with tilt angles from ``L'' %model are usually better in agreement with experimental data with respect to the \textit{diagonal %approximation} approximation
and, finally, iii) the minimum difference with the experimental data occurs when the enhancement of the diffusion tensor along the polar direction ($K_{\perp \theta}$) [Eq.~(\ref{enhaced_K})] is taken into account (Table~\ref{table::min}, see first column of the left-hand side).~In addition, the results obtained for a heliospheric region where solar latitudes are lower than $|30^\circ|$ exhibit a behavior similar to those lower than $|5.7^\circ|$, but with values of $\eta_{\rm RMS}$ (in percentage) larger by about several percents.~In Figs.~\ref{Fig:Bess_1997}--\ref{Fig:pamela}, the differential intensities determined with HelMod code are shown and compared to the experimental data of BESS--1997, AMS--1998, BESS--1998 and PAMELA--2006/08, respectively; in the same figures, the dashed line is the LIS [Eqs.~(\ref{BHH_LIS},~\ref{J_HelMod})] with normalization constants $J_0$ treated in Sect.\ref{LIS_discussion}.~These modulated intensities are the ones calculated for a heliospheric region where solar latitudes are lower than $|5.7^\circ|$, using the enhancement of the diffusion tensor along the polar direction ($K_{\perp \theta}$) [Eq.~(\ref{enhaced_K})] and including particle drift effects with the values of tilt angle from ``L'' model.
\par
Finally, it has be concluded that the present code combining diffusion and drift mechanisms is also suited to describe the modulation effect in periods when the solar activity is no longer at the maximum.~%Furthermore - as already %remarked -, the experimental data from observations with high statistics over a long duration (like those from the AMS-02 spectrometer) will allow one to undertake a systematic investigation of the solar modulation effect over a %period larger than a solar cycle.
\subsection{Dependence on the Extension of Heliosphere}
\label{Effective_Hel}
In Sects.~\ref{H_A_Results} and~\ref{R_L_Tilt_Angle}, the simulated differential intensities were obtained from a LIS [described by Eqs.~(\ref{BHH_LIS},~\ref{J_HelMod})] propagating through a spherical heliosphere with a radius of 100\,AU down to Earth.~However, the physical dimensions of the heliosphere also depends on the speed of solar wind% which, in turn, is related to solar %activity
.~In HelMod code, the simulated modulated intensities are determined by the properties of the diffusion tensor (Sects.~\ref{Model},~\ref{FFModel},~\ref{antysymmetric_part} and Appendix~\ref{app1}), whose elements are related to the actual value of the diffusion parameter.~%The diffusion parameter is a function of the intensity of solar activity and varies %with solar polarity and phase (Sects.~\ref{FFModel},~\ref{dynamic}).~
$K_0$ acts as a \textit{scaling factor} for the overall modulation effect.~It was indirectly determined from neutron monitor measurements, thus, it is expected to be sensitive to the overall modulation effect (from the heliosphere boundary down to Earth), but almost independent of the variation of heliosphere dimensions.
\par
The radial distance of the heliosphere was varied from 80 up 120\,AU.~The corresponding simulated differential intensities were compared to the experimental data from BESS--2002 %~\citep{BESS_Astropart}
[data collected during high solar activity (Sect.~\ref{H_A_Results})] and PAMELA--2006/08 %~\citep{Pamela_2011}
[data collected when the solar activity was no longer large (Sect.~\ref{R_L_Tilt_Angle})],~i.e., when heliosphere is expected to be smaller or larger (and possibly no longer spherical) than 100\,AU, respectively.~
\par
The values of $\eta_{\rm RMS}$ [Eq.~(\ref{eta_rms})] in percentage calculated for spherical heliospheres with radii of 80, 90, 110 and 120\,AU are shown in Table~\ref{tab::distHB} and compared with those calculated with a radius of 100\,AU (see Tables~\ref{table::max_noen},~\ref{table::min}).~For BESS--2002, the simulated intensities were obtained i) using the ``L'' model for the tilt angle ($\alpha_{\rm t}$) [Sect.~\ref{dynamic} and~\citep{Hoek95,wsoWeb}], ii) with  $K_{\perp \theta}=K_{\perp r}$ independently of the latitude and iii) inside a heliospheric region where solar latitudes are lower than $|30^\circ|$.~For PAMELA--2006/08, the simulated intensities were obtained a) using the ``L'' model for the tilt angle ($\alpha_{\rm t}$) [Sect.~\ref{dynamic} and~\citep{Hoek95,wsoWeb}], b) with an enhancement of the diffusion tensor along the polar direction ($K_{\perp \theta}$) [Eq.~(\ref{enhaced_K})] and c) inside a heliospheric region where solar latitudes are lower than $|5.7^\circ|$.~From inspection of Table~\ref{tab::distHB}, one can remark that, within 2.3\%, the simulated differential intensities for spherical heliospheres with radii of 80, 90 and 110\,AU are compatible with that with a radius of 100\,AU; slightly larger values of $\eta_{\rm RMS}$ were obtained for a spherical heliosphere with a radius of 120\,AU.
%%%%%%%%%%%%%%%%%%%%%%%%%%%%%%%%%%%
\begin{table}%[t]
\begin{center}
\begin{tabular}[t]{ccccccc}
      &     & 80\,AU & 90\,AU & 100\,AU &110\,AU & 120\,AU   \\
\hline
 BESS-2002  &  & 14.5  & 12.00 & 12.2 & 13.0 & 14.5  \\
 \hline
PAMELA--2006/08 &     & 5.7 & 6.1  &  7.1  & 7.7 & 10.8 \\
\hline
\end{tabular}
\caption{$\eta_{\rm RMS}$ (in percentage) calculated for a spherical heliosphere with radius of 80, 90, 110 and 120\,AU and compared with those obtained with a radius of 100\,AU: for BESS--2002 (see Table~\ref{table::max_noen}), for PAMELA--2006/08 (see Table~\ref{table::min}).}\label{tab::distHB}
\end{center}
\end{table}
%%%%%%%%%%%%%%%%%%%%%%%%%%%%%%%%%%%
\par
The sensitivity of this approach was estimated from the differences of the simulated intensities with radii of 80, 90, 110 and 120\,AU with that with a radius of 100\,AU for protons with energies above 30\,GeV,~i.e., for an energy region in which the spectrum is unaffected by modulation and, thus, no difference is expected.~For this purpose, we defined the quantity [see also Eqs.~(\ref{eta_rms},~\ref{eta_i})]
\begin{equation}\label{eta_rms_hat}
 \hat{\eta}_{\rm RMS,h}=\sqrt{\frac{\sum_i \left({\hat{\eta}_{i, \rm h}}/\sigma_{\hat{\eta},i, \rm h}\right)^2}{\sum_i 1/\sigma^2_{\hat{\eta},i, \rm h}} }
\end{equation}
with
\begin{equation}\label{eta_i_hat}
 \hat{\eta}_{i, \rm h}=\frac{ f_{\rm h}(T_i) -f_{\rm 100AU}(T_i) }{f_{\rm 100AU}(T_i)},
\end{equation}
where $f_{\rm h}(T_i)$ is the differential intensity of $i$-th energy bin (above 30\,GeV), $\sigma_{\hat{\eta},i, \rm h}$ is the error due to Monte Carlo uncertainties for $i$-th energy bin, $f_{\rm 100AU}(T_i)$ is the differential intensity computed with a radius of 100\,AU and, finally ``h'' indicates 80, 90, 110 and 120\,AU.~$\hat{\eta}_{\rm RMS,h}$ resulted equal to about  2.3\% for heliospheres with 80, 90, 110 and 120\,AU.~Thus, the modulated intensities for heliospheres with radii of 80, 90, 100, 110\,AU (and also 120\,AU for BESS-2002) are in agreement among them and experimental data within the present sensitivity of about 2.3\% of the current approach; at 120\,AU the simulated intensity is marginally non compatible with that obtained with 100\,AU for PAMELA--2006/08.~These results indicated that, as expected, the diffusion parameter %- which, as already mentioned, is a function of the intensity of solar activity and varies with solar %polarity and phase (Sects.~\ref{FFModel},~\ref{dynamic}) -
almost accounts for effects related to the variation of the physical dimensions of the heliosphere within the present approximations.~%Finally, as already noted, the high precision data from AMS-02 will allow one to achieve a %higher sensitivity and, possibly, a better approximated treatment of the actual heliosphere dimension as a function of the solar activity.
\subsection{Dependence on Heliospheric Latitude}
\label{Heliosph_latid_dep}
Observations made by the Ulysses spacecraft~\citep{simpson1992} in the inner
heliosphere could determine a latitudinal dependence of GCR (mostly protons) intensity with an equatorial southward offset minimum and a North polar excess.~This dependence was discussed also in terms of modulation models which were including particle drift effects~\citep[e.g., see][]{simpson1996,Ulysses}.~For protons with energies larger than 100\,MeV, \citet{simpson1996b} expressed their results in terms of the solar latitude and found that i) the latitudinal gradient is $\approx (0.33 \pm 0.02)\% \,\deg^{-1}$, ii) the counting rate minimum is nearly constant in a latitudinal region of $\approx -(15^{\circ}$--$5^{\circ})$ [Figure~2 of~\citet{simpson1996b}] at $\approx 1.35\,$AU [e.g., see~\citep{simpson1996,Ulysses}] and iii) the rate at the minimum is about $\approx 80$\% with respect to that at $\approx 80^{\circ}$.~\citet{Ferr_2003} have shown that the latitudinal dip - found with the Ulysses fast scan - can be reproduced in a model using the Parker standard field and a polar enhancement of the diffusion tensor.~
%%%%%%%%%%%%%%%%%%%%%%%%%%%%%%%%%%%%%%%%%%%%
\begin{figure}[t]
\centering
\includegraphics[width=0.4\textwidth]{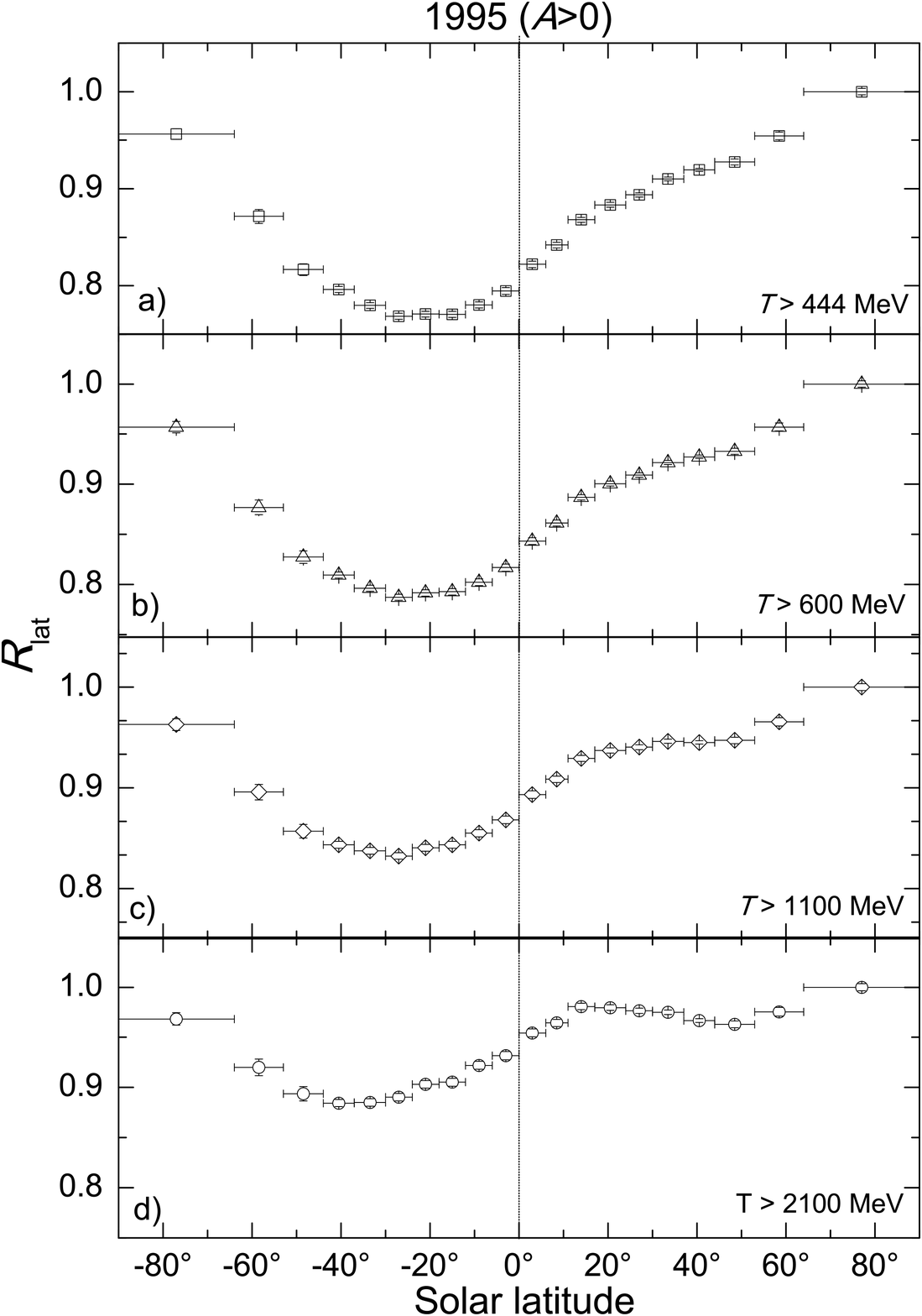}
\includegraphics[width=0.4\textwidth]{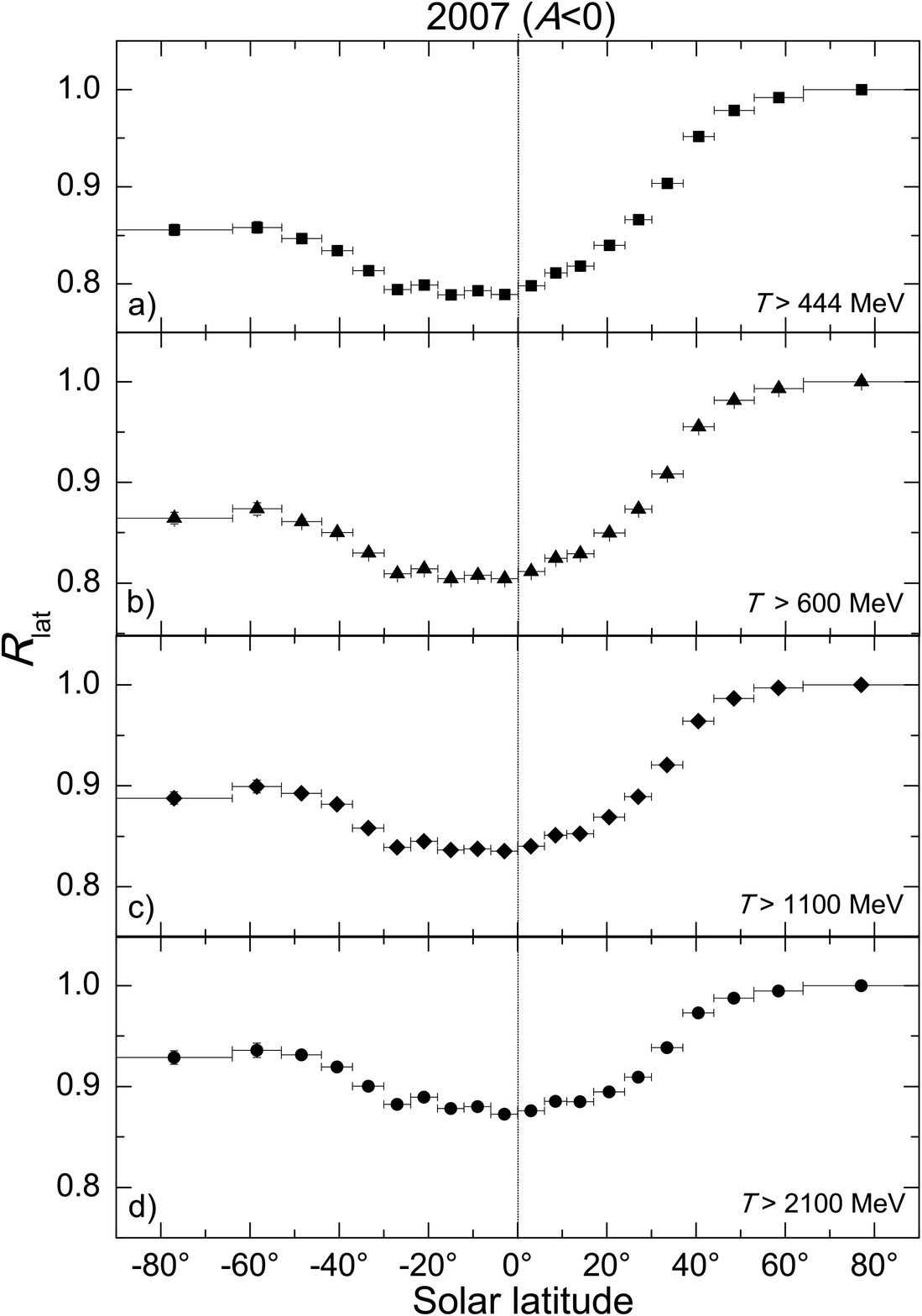}
\caption{$R_{\rm lat}$ calculated at 1\,AU as a function function of solar latitude for a) $T_e > 444$\,MeV, b) $T_e > 600$\,MeV, c) $T_e > 1200 $\,MeV and d) $T_e > 2100 $\,MeV: left-hand side during 1995 ($A>0$), right-hand during 2007 ($A<0$).}
\label{fig_simlat}
\end{figure}
%%%%%%%%%%%%%%%%%%%%%%%%%%%%%%%%%%%%%%%
\par
Using HelMod code, we could investigate the latitudinal dependence of the differential intensity at 1\,AU, above 444\,MeV (as so far treated) up to 200\,GeV.~The heliosphere was subdivided in 20 regions equally spaced with respect to the co-latitudinal parameter $\mu(\theta)$ [Sect.~\ref{Code}].~The total fluxes obtained in each region were divided by the maximum flux occurring at the North pole, thus, $R_{\rm lat}$ represents the normalized flux (to that at the North pole) as a function of the co-latitude.~In addition, the values of $R_{\rm lat}$ were calculated for periods of opposite magnetic polarities and compatible with Ulysses pole-to-pole fast scans,~i.e., for the years 1995 with $A\,>\,0$ and 2007 with $A\,<\,0$.~$R_{\rm lat}$ can be equivalently expressed as a function of the solar latitude for a comparison with the results obtained by \citet{simpson1996b}.~$R_{\rm lat}$ as a function the solar latitude is shown in Fig.~\ref{fig_simlat} for the year 1995 (left-hand side) and 2007 (right-hand side).~$R_{\rm lat}$ is also shown as a function of the minimum kinetic energy accounted for protons ($T_e$),~i.e., a) $T_e > 444$\,MeV, b) $T_e > 600$\,MeV, c) $T_e > 1100 $\,MeV and d) $T_e > 2100 $\,MeV.~By inspection of Fig.~\ref{fig_simlat}, for the year 1995 one can remark that $R_{\rm lat}$ has 1) a latitudinal gradient of $(0.23 \pm 0.01)\%\,\deg^{-1}$, 2) an equatorial southward offset minimum in the latitudinal region $\approx -(18^{\circ}$--$5^{\circ})$ with $T_e > 444$\,MeV and 3) at this minimum, the flux is about $\approx 80$\% of that at the North pole.~Thus, the simulated fluxes reproduce the features of the experimental data (above 100\,MeV) exhibited in Figures~2 and~3 of~\citet{simpson1996b} [see also~\citep{heber2008}] regarding the period of 1995 Ulysses fast scan.~However with increasing $T_e$, the latitudinal gradient decreases (Fig.~\ref{fig_simlat}, left-hand side).~In the year 2007, a similar minimum is exhibited for $T_e > 444$\,MeV with a $\approx 10$\% North--South poles asymmetry; with increasing $T_e$, this asymmetry gradually disappears and the flux reduction on the equatorial region is less pronounced down to $\approx 88$\% with $T_e > 2100$\,MeV (Fig.~\ref{fig_simlat}, right-hand side).
\par
It is worthwhile to note that in HelMod code the magnetic field structure is treated similarly in North and South hemisphere approximating Parker's magnetic-field with that suggested
by~\cite{JokipiiKota89} (Sect.~\ref{HMF}).~However, the current 2-D model uses the complete $2 \times 2$ diffusion tensor (see Sects.~\ref{HMF},~\ref{Code} and Appendix~\ref{app1}) which contains both symmetric and antisymmetric components in the off-diagonal terms.~The symmetric component of the off-diagonal terms [Eq.~(\ref{tens_our5})] is determined by the divergence-free IMF used, which exhibits a latitudinal component arising from
the modification by~\citet{JokipiiKota89} [Eqs.~(\ref{FUN_ParkerField},~\ref{lat_comp},~\ref{delta_sin})].~In the framework of the present 2-D model, the North polar excess and equatorial southward offset minimum shown in Fig.~\ref{fig_simlat} originate from the non-zero symmetric component of the off-diagonal terms.~The actual extension of the dip is related to both the enhancement of the diffusion tensor in the polar regions [Eq.~(\ref{enhaced_K})] and drift effects.
\section{Conclusions}
\label{Cncl_sect}
A systematic investigation of the solar modulation effect on the propagation of cosmic protons through the heliosphere down to the Earth was carried out comparing experimental observations performed during the solar cycle 23 and simulated differential intensities obtained using HelMod code.~The simulated spectra were derived from a LIS [Eqs.~(\ref{BHH_LIS},~\ref{J_HelMod})], whose normalization constant ($J_0$) depends on the experimental set of data (see discussion in Sect.\ref{LIS_discussion}).~The stochastic 2D Monte Carlo (HelMod) code includes i) a fully treated diffusion tensor with symmetric and antisymmetric off-diagonal elements, b) a diffusion parameter which is a function of the intensity of solar activity and varies with solar polarity and phase (Sects.~\ref{FFModel},~\ref{dynamic}) and c) a magnetic-field which is modified with respect to Parker's magnetic field in the polar region as proposed by~\cite{JokipiiKota89} (Sect.~\ref{HMF}).~
\par
For observations performed during high solar activity, the simulated intensities (found with a better agreement to experimental data) were obtained i) using the ``L'' model for the tilt angle ($\alpha_{\rm t}$) [Sect.~\ref{dynamic} and~\citep{Hoek95,wsoWeb}], ii) with  $K_{\perp \theta}=K_{\perp r}$ independently of the latitude and iii) inside a heliospheric region where solar latitudes are lower than $|30^\circ|$.~For observations performed when solar activity is no longer at the maximum, the simulated intensities (found with a better agreement to experimental data) were obtained a) using the ``L'' model for the tilt angle ($\alpha_{\rm t}$) [Sect.~\ref{dynamic} and~\citep{Hoek95,wsoWeb}], b) with an enhancement of the diffusion tensor along the polar direction ($K_{\perp \theta}$) [Eq.~(\ref{enhaced_K})] and c) inside a heliospheric region where solar latitudes are lower than $|5.7^\circ|$.~
\par
In addition (within 2.3\%), the simulated differential intensities for spherical heliospheres with radii of 80, 90 and 110\,AU (and also 120\,AU for BESS-2002) are compatible with that with a radius of 100\,AU; a slightly lower agreement was obtained for a spherical heliosphere with a radius of 120\,AU for PAMELA--2006/2008.~These results indicated that, within the present approximations, the diffusion parameter almost accounts for effects related to the variation of the physical dimensions of the heliosphere.
\par
The simulated modulated spectrum determined for the year 1995 exhibits a latitudinal gradient of $(0.23 \pm 0.01)\% \, \deg^{-1}$, an equatorial southward offset minimum in the latitudinal region $\approx -(18^{\circ}$--$5^{\circ})$ with $T_e > 444$\,MeV and at this minimum the flux is about $\approx 80$\% of that at North pole.~Thus, the simulated fluxes reproduce the features of the experimental data from Ulysses fast scan in 1995~\citep{simpson1996b}.
\par
Although the treatment is highly simplified with respect to the complexity of physical mechanisms responsible for modulation effects, the overall satisfactory agreement found allows one to remark that the choice of parameters regarding the structure of IMF, diffusion tensor, diffusion parameter and tilt angle is almost appropriate to describe the experimental data.~Finally, the experimental data from accurate observations over a long duration (like those from the AMS-02 spectrometer) will allow one to undertake a deeper systematic investigation of solar modulation effects over a period longer than a solar cycle.~Thus, possibly, further advancements can be put forward in the present approximations on the transport of GCR's through the heliosphere, for instance those at low rigidities,
the spatial and rigidity properties of diffusion tensor.
% the dependence of diffusion tensor elements on the rigidity and solar colatitude as a function of the distance from Sun and the properties of off-diagonal elements of the diffusion tensor.
%
\appendix
\section*{Appendix}
\section{Diffusion Tensor and Stochastic Differential Equations}
\label{app1}
In a reference frame with the 3rd coordinates along the average magnetic field, the matrix of the diffusion tensor used in Eqs.~(\ref{eq_parker},~\ref{eq_parker1})
is given by~\citep[e.g., see][]{Jokipii1971}:
\[
 K_{ik}=\left|\begin{array}{lll}
         K_{\perp r}&  -K_A		& 0\\
	 K_A	& K_{\perp \theta} 	&0\\
	 0	& 0		&   K_{||}
        \end{array}\right|.
        \]
In heliocentric spherical coordinates $(r,\theta,\phi)$, for instance those used in Eqs.~(\ref{eq::app1}--\ref{eq::app3}), the matrix elements of the $3 \times 3$ tensor are found, for instance, in Equation~(17) of~\citet{burg2008}.~In a 2-D approximation, the matrix elements of the resulting tensor are:% reduced to:($2 \times 2$)
\begin{eqnarray}
 &&K_{rr}=K_{\perp \theta}\, \sin^2\xi + \cos^2\xi \left(K_{||}\,\cos^2\psi + K_{\perp r}\,\sin^2\psi\right)\label{tens_generic1},\\
 &&K_{\theta\theta}= K_{\perp \theta}\, \cos^2\xi + \sin^2\xi \left(K_{||}\,\cos^2\psi + K_{\perp r}\,\sin^2\psi\right),\\
 &&K_{r\theta}= - K_A \sin\psi+ \sin\xi\cos\xi \left(K_{||}\cos^2\psi + K_{\perp r}\,\sin^2\psi - K_{\perp \theta}\right),\\
 &&K_{\theta r}=  K_A \sin\psi+\sin\xi\cos\xi \left(K_{||}\cos^2\psi + K_{\perp r}\,\sin^2\psi - K_{\perp \theta}\right)\label{tens_generic4}
\end{eqnarray}
with $\tan\psi=-{B_\phi}/({\sqrt{B^2_r+B^2_\theta})}$ and $\tan\xi={B_\theta}/{B_r}$ [see Figure~6 of~\citet{burg2008}], where $\psi$ is the spiral angle (for a standard Parker IMF $\tan\psi$ reduces to Eq.~(\ref{spiral_angle})
and $\tan\xi=0$), and with
\begin{eqnarray}
 \label{A_kparr} && K_{||} = \beta \, k_1(r,t) \,K_{P}(P,t)\,\left[\frac{B_{\oplus}}{3B}\right] ,\\
 \label{A_kperp} && K_{\perp r} = \rho_k K_{||},\\
 \label{A_kperpT} && K_{\perp \theta} =\iota(\theta)\rho_k K_{||},
\end{eqnarray}
where $\rho_k=0.05$ and $\iota(\theta)$ is a step function that is 1 in equatorial region and 10 in polar region (Sect.~\ref{Model}).~The diffusion parameter  $k_1(r,t)$ is replaced by $K_0$ for an effective heliosphere of 100\,AU (Sect.~\ref{FFModel}).~Furthermore, the matrix elements of the later tensor consist of a symmetric ($K^S_{ik}$) and antisymmetric ($K^A_{ik}$) part:
\begin{equation}
\label{A_split} K_{ik}=K^A_{ik}+K^S_{ik}
\end{equation}
with the antisymmetric part related to drift velocity [Eq.~(\ref{v_drift})], $ \vec{v}_{\rm d}$, and treated in Sect.~\ref{antysymmetric_part}.~Finally, using Eqs.~(\ref{A_kparr}--\ref{A_split}), Eqs.~(\ref{tens_generic1}--\ref{tens_generic4}) can be re-written as:%:we re-write as:
\begin{eqnarray}
\label{tens_our1} && K_{rr}=K^S_{rr}= \beta  K_0 \,K_{P}(P,t)\!\left[\frac{B_{\oplus}}{3B}\right] \! \left[
          \iota(\theta)\rho_k\,\sin^2\xi + \cos^2\xi\!\left(\cos^2\psi +\rho_k\,\sin^2\psi\right) \right],\\
\label{tens_our2} && K_{\theta\theta}=K^S_{\theta\theta}=  \beta  K_0 \,K_{P}(P,t)\!\left[\frac{B_{\oplus}}{3B}\right] \!\left[
           \iota(\theta)\rho_k\, \cos^2\xi + \sin^2\xi\!\left(\,\cos^2\psi + \rho_k\,\sin^2\psi\right) \right],\\
\label{tens_our3} && K_{r\theta}=  K^A_{r\theta}+ K^S_{r\theta},\\
\label{tens_our4} && K_{\theta r}=  K^A_{\theta r}+K^S_{r\theta}
 \end{eqnarray}
 with
 \begin{eqnarray}
\label{tens_our5} && K^S_{r\theta}=\beta \, K_0 \,K_{P}(P,t)\,\left[\frac{B_{\oplus}}{3B}\right]  \left\{
            \sin\xi\cos\xi\left[\cos^2\psi + \rho_k\,\sin^2\psi - \iota(\theta)\rho_k\right]\right\},\\
\label{tens_our6} &&  K^A_{\theta r}=-K^A_{r\theta}= K_A \sin\psi.
\end{eqnarray}
\par
Equations~(\ref{eq_parker},~\ref{eq_parker1}) can be re-expressed in heliocentric spherical coordinates as:
\begin{eqnarray}\label{eq::ParKerEQ_sph}
 \frac{\partial U}{\partial t}&=&   \frac{1}{r^2}\frac{\partial  } {\partial r}\left(r^2 K^S_{rr}\frac{\partial}{\partial r} U + r K^S_{r\theta}\frac{\partial}{\partial \theta} U + \frac{r}{\sin\theta} K^S_{r\phi}\frac{\partial}{\partial \phi} U\right)  \nonumber \\
&&  + \frac{1}{r \sin\theta}\frac{\partial  }{\partial \theta}\left(\sin\theta K^S_{\theta r}\frac{\partial}{\partial r} U + \frac{\sin\theta}{r} K^S_{\theta \theta}\frac{\partial}{\partial \theta} U + \frac{1}{r^2} K^S_{\theta \phi}\frac{\partial}{\partial \phi} U \right)  \nonumber \\
&&  + \frac{1}{r \sin\theta}\frac{\partial  }{\partial \phi}\left( K^S_{\phi r}\frac{\partial}{\partial r} U + \frac{1}{r} K^S_{\phi \theta}\frac{\partial}{\partial \theta} U + \frac{1}{r \sin\theta} K^S_{\phi \phi}\frac{\partial}{\partial \phi} U\right) \nonumber \\
&&  - \frac{1}{r^2}\frac{\partial r^2 V_r U}{\partial r} -  \frac{1}{r \sin\theta}\frac{\partial  \sin\theta V_\theta U}{\partial \theta} - \frac{1}{r \sin\theta}\frac{\partial   V_\phi U}{\partial \phi} \nonumber \\
&&  - \frac{1}{r^2}\frac{\partial r^2  v_{d,r} U}{\partial r} -  \frac{1}{r \sin\theta}\frac{\partial  \sin\theta  v_{d,\theta} U}{\partial \theta} - \frac{1}{r \sin\theta}\frac{\partial  v_{d,\phi} U}{\partial \phi}  \nonumber \\
&&  + \frac{1}{3}\left( \frac{1}{r^2}\frac{\partial r^2 V_r}{\partial r} +  \frac{1}{r \sin\theta}\frac{\partial  \sin\theta V_\theta}{\partial \theta} + \frac{1}{r \sin\theta}\frac{\partial   V_\phi}{\partial \phi} \right)\frac{\partial}{\partial T}(\alpha_{\rm rel} T U),
\end{eqnarray}
where $U$ is the number density of GCRs (Sect.~\ref{Model}) and $T$ is the kinetic energy.~In turn, in a 2-D (radial distance and co-latitude) approximation, Eq.~(\ref{eq::ParKerEQ_sph}) can be re-expressed as:
\begin{eqnarray}\label{eq::ParKerEQ_sph_2D}
 \frac{\partial U}{\partial t}&=&   \frac{1}{r^2}\frac{\partial  } {\partial r}\left(r^2 K^S_{rr}\frac{\partial}{\partial r} U + r K^S_{r\theta}\frac{\partial}{\partial \theta} U \right)   + \frac{1}{r \sin\theta}\frac{\partial  }{\partial \theta}\left(\sin\theta K^S_{\theta r}\frac{\partial}{\partial r} U + \frac{\sin\theta}{r} K^S_{\theta \theta}\frac{\partial}{\partial \theta} U\right)  \nonumber \\
&&  - \frac{1}{r^2}\frac{\partial r^2 V_r U}{\partial r} -  \frac{1}{r \sin\theta}\frac{\partial  \sin\theta V_\theta U}{\partial \theta}  - \frac{1}{r^2}\frac{\partial r^2  v_{d,r} U}{\partial r} -  \frac{1}{r \sin\theta}\frac{\partial  \sin\theta  v_{d,\theta} U}{\partial \theta}  \nonumber \\
&&  + \frac{1}{3}\left( \frac{1}{r^2}\frac{\partial r^2 V_r}{\partial r} +  \frac{1}{r \sin\theta}\frac{\partial  \sin\theta V_\theta}{\partial \theta} \right)\frac{\partial}{\partial T}(\alpha_{\rm rel} T U).
\end{eqnarray}
\par
Let us define the variable $\mu=\cos(\theta)$, then we obtain
\begin{equation}\label{eq::partheta_mu}
\partial \theta = -(1-\mu^2)^{-0.5}\partial \mu.
\end{equation}
In addition, one can introduce the function
\begin{equation}\label{eq::F_mu}
F=U\,r^2.
\end{equation}
Using Eqs.~(\ref{eq::ParKerEQ_sph_2D}--\ref{eq::F_mu}), for a SW radially propagating  (i.e. $V_{{\rm sw},r}=V_{\rm sw}$)  we find:
\begin{eqnarray}\label{eq::ParkerEQ_xSDE}
 \frac{\partial F}{\partial t}&=&-\frac{\partial}{\partial r}\left[\frac{F}{r^2}\frac{\partial}{\partial r}(r^2 K^S_{rr}) \right]
                                 -\frac{\partial}{\partial r}\left[-F \frac{\partial}{\partial \mu}\left(\frac{ K^S_{r\theta}\sqrt{1-\mu^2}}{r}\right)   \right]
				 -\frac{\partial}{\partial r}\left[F(V_{\rm{sw}}+v_{d,r}  )\right]
\nonumber\\
			      && -\frac{\partial}{\partial \mu}\left[-\frac{F}{r^2} \frac{\partial}{\partial r}\left(r K^S_{\theta r}\sqrt{1-\mu^2}\right)   \right]
                                 -\frac{\partial}{\partial \mu}\left\{F \frac{\partial}{\partial \mu}\left[ \frac{K^S_{\theta \theta}(1-\mu^2)}{r^2}\right] \right\}
\nonumber\\
				 &&-\frac{\partial}{\partial \mu}\left[-F\frac{ v_{d,\theta}\sqrt{1-\mu^2}}{r} \right]
			       -\frac{\partial}{\partial T}\left[-F\frac{\alpha_{\rm rel} T}{3 r^2}\frac{\partial V_{\rm sw}r^2}{\partial r}     \right]
\nonumber\\
				 &&+\frac{1}{2}\frac{\partial}{\partial r}\frac{\partial}{\partial r}[2K^S_{rr}F]
                                 +\frac{1}{2}\frac{\partial}{\partial r}\frac{\partial}{\partial \mu}\left[- \frac{2 K^S_{r\theta}\sqrt{1-\mu^2}}{r}   F\right]
\nonumber\\
                              &&+\frac{1}{2}\frac{\partial}{\partial \mu}\frac{\partial}{\partial r}\left[- \frac{2 K^S_{\theta r}\sqrt{1-\mu^2}}{r}   F\right]
			        +\frac{1}{2}\frac{\partial}{\partial \mu}\frac{\partial}{\partial \mu}\left[ \frac{2 K^S_{\theta \theta}(1-\mu^2)}{r^2} F\right].
\end{eqnarray}
Furthermore, following the treatment discussed in Sections~4.3--4.3.5 of~\citep{gardiner1985}, one can a) express the Fokker-Plank equation involving $F$ - which, in turn, is a function of $\mathbf q=(r,\mu, T)$ - as:
\begin{equation} \label{eq::fokkerGeneral}
 \frac{\partial}{\partial t} F = - \sum_i\frac{\partial}{\partial q_i}[A_i(\mathbf q, t )F] + \frac{1}{2}\sum_{i,j} \frac{\partial}{\partial q_i} \frac{\partial}{\partial q_j} \{[\mathbf{\tilde D}(\mathbf q, t)]_{ij}\,F \}
\end{equation}
with $\mathbf{\tilde D=\tilde L\tilde L^T}$ and b) obtain the equivalent set of differential equations
\begin{equation}\label{eq::ito_SDE}
{\rm d}\mathbf q = \mathbf A (\mathbf q,t) {\rm d}t + \mathbf{\tilde L}(\mathbf q, t) \,{\rm d}\mathbf W (t),
\end{equation}
where
$\mathbf A (\mathbf q,t)\, {\rm d}t$ accounts for the so-called \textit{advective processes}~\cite[e.g.,][]{kruell1994},
$\mathbf{\tilde L}(\mathbf q, t)\, {\rm d}\mathbf W (t)$ is the \textit{stochastic term} containing ${\rm d}\mathbf W (t)$ which is the increment of the so-called \textit{Wiener process}~\citep[e.g., Section~4.3 of][]{gardiner1985}.~Equations~(\ref{eq::ito_SDE}) are termed \textit{stochastic differential equations} (SDEs).
\par
Furthermore, one can note that i) the first right-hand term of Eq.~(\ref{eq::fokkerGeneral}) is equal to those included in the first three lines of Eq.~(\ref{eq::ParkerEQ_xSDE}) and ii) the second right-hand term of Eq.~(\ref{eq::fokkerGeneral}) is equal to those included in the fourth and fifth line of Eq.~(\ref{eq::ParkerEQ_xSDE}).~Thus,~using Eqs.~(\ref{eq::ParkerEQ_xSDE},~\ref{eq::fokkerGeneral}) one derives:
\begin{eqnarray}\label{eq::SDE_A_vector}
 \mathbf A = \left[\begin{array}{c}
              \frac{1}{r^2}\frac{\partial}{\partial r}(r^2 K^S_{rr}) - \frac{\partial}{\partial \mu}\left(\frac{ K^S_{r\theta}\sqrt{1-\mu^2}}{r}\right) + V_{\rm{sw}}+v_{d,r}\\
             -\frac{1}{r^2} \frac{\partial}{\partial r}\left(r K^S_{\theta r}\sqrt{1-\mu^2}\right)+ \frac{\partial}{\partial \mu}\left[ \frac{K^S_{\theta \theta}(1-\mu^2)}{r^2}\right]-\frac{v_{d,\theta}\sqrt{1-\mu^2} }{r}\\
              -\frac{\alpha_{\rm rel} T}{3 r^2}\frac{\partial V_{\rm sw}r^2}{\partial r}
                   \end{array}\right]
\end{eqnarray}
and
\begin{eqnarray}\label{eq::SDE_D_matrix}
 \mathbf {\tilde D}= \left[\begin{array}{cc}
                            2K^S_{rr}					& - \frac{2 K^S_{r\theta}\sqrt{1-\mu^2}}{r}  \\
			    - \frac{2 K^S_{\theta r}\sqrt{1-\mu^2}}{r} 	& \frac{2 K^S_{\theta \theta}(1-\mu^2)}{r^2}
                           \end{array} \right].
\end{eqnarray}
As discussed by~\citet*{Pei2010} - see Section 2 therein -, the matrix $\mathbf {\tilde D}$ can be downgraded to a two-by-two matrix, because second order acceleration mechanisms are not considered in Eqs.~(\ref{eq_parker},~\ref{eq_parker1}).
\par
As already shown by~\citet*{Pei2010} - see Appendix B therein -, the matrix $\mathbf{\tilde L}$ is not unique.~However, %, since it is the square root of $\mathbf {\tilde D}$;
there is only one positive definite,~i.e.,
\begin{equation}\label{eq::matrixL}
  \mathbf {\tilde L}= \left[\begin{array}{cc}
                            \left[\frac{K^S_{rr}K^S_{\theta\theta}-(K^S_{r\theta})^2}{0.5 K^S_{\theta\theta}}\right]^{1/2}	 &  - K^S_{r\theta}\left(\frac{2}{K^S_{\theta\theta}}\right)^{1/2} \\
			    0 	&  \left[\frac{2 K^S_{\theta \theta}(1-\mu^2)}{r^2} \right]^{1/2}
                           \end{array} \right].
\end{equation}
Finally, for a 2-dimensional model (like that treated in Sects.~\ref{HMF},~\ref{antysymmetric_part}), from Eqs.~(\ref{eq::ito_SDE},~\ref{eq::SDE_A_vector}) and using Eq.~(\ref{eq::matrixL}) one finds the following SDEs:
\begin{eqnarray}
\label{eq::app1} {\rm d}r  &=& \left[\frac{1}{r^2}\frac{\partial}{\partial r}(r^2 K^S_{rr}) - \frac{\partial}{\partial \mu}\left(\frac{ K^S_{r\theta}\sqrt{1-\mu^2}}{r}\right) + V_{\rm{sw}}+v_{{\rm d},r}\right] {\rm d} t +\nonumber\\
           &&    + \sqrt{\frac{K^S_{rr}K^S_{\theta\theta}-(K^S_{r\theta})^2}{0.5 K^S_{\theta\theta}}}\, {\rm d}W_r   -  K^S_{r\theta}\sqrt{\frac{2}{K^S_{\theta\theta}}\,}\, {\rm d}W_\mu,
            \\
\label{eq::app2} {\rm d} \mu&=& \left\{-\frac{1}{r^2} \frac{\partial}{\partial r}\left(r K^S_{\theta r}\sqrt{1-\mu^2}\right)+ \frac{\partial}{\partial \mu}\left[ \frac{K^S_{\theta \theta}(1-\mu^2)}{r^2}\right]-\frac{v_{{\rm d},\theta}\sqrt{1-\mu^2}}{r}\right\} {\rm d} t +\nonumber\\
           &&  +  \sqrt{\frac{2 K^S_{\theta \theta}(1-\mu^2)}{r^2} \,}\, {\rm d}W_\mu, \\
\label{eq::app3} {\rm d} T  &=&  - \frac{2}{3} \frac{\alpha_{\rm rel} V_{\rm sw} T}{r} {\rm d} t
\end{eqnarray}
with ${\rm d}W_i$ [$i =r, \mu(\theta)$] the increment of the Wiener process.~It has be remarked that the above usage of Eq.~(\ref{eq::matrixL}) ensures imaginary terms do not appear in Eqs.~(\ref{eq::app1}--\ref{eq::app6}).~For an IMF described by a standard Parker field [Eq.~(\ref{FUN_ParkerField})] requiring $K^S_{r\theta}=0$, the above SDEs reduce to
\begin{eqnarray}
\label{eq::app4} {\rm d} r  &=&  \frac{1}{r^2}  \left[\frac{\partial}{\partial r}\left( r^2 K^S_{rr}\right)\right]  {\rm d} t + \left( V_{\rm sw}+ v_{{\rm d},r}   \right) {\rm d} t + \sqrt{2K^S_{rr}\, }\, {\rm d}W_r,  \\
\label{eq::app5} {\rm d} \mu &=& \frac{1}{r^2} \left\{\frac{\partial}{\partial \mu}\left[ (1-\mu^2)K^S_{\theta\theta}\right]\right\}  {\rm d} t - \frac{v_{{\rm d},\theta} \,\sqrt{1-\mu^2}}{r}\,   {\rm d} t \nonumber\\
                                       & & + \sqrt{ \frac{2\,K^S_{\theta \theta}\, (1-\mu^2)}{r^2} \,} \,{\rm d}W_\mu, \\
\label{eq::app6} {\rm d} T  &=&  - \frac{2}{3} \frac{\alpha_{\rm rel} V_{\rm sw} T}{r}  {\rm d} t.
\end{eqnarray}
\par
Gardiner~(1985) - see Section~4.3.1 therein - demonstrated that, following a \textit{Euler--Cauchy procedure}, the SDEs can be approximated in terms of the increments $\Delta r$, $\Delta \mu$ and $\Delta T$ occurring after a time  $\Delta t$ has elapsed.
The corresponding increment of the Wiener Process  is given by $\omega_i\sqrt{\Delta t}$ [with $i =r, \mu(\theta)$], where $\omega_i$ is a random number following a Gaussian distribution with a mean of zero and a standard deviation of one~(e.g.,~see \citealt{kruell1994} and appendix A of~\citealt{Pei2010}).~As a consequence, the SDEs [Eqs.~(\ref{eq::app1}--\ref{eq::app3})] can be approximated by:
\begin{eqnarray}
\Delta r  &=& \left[\frac{1}{r^2}\frac{\partial}{\partial r}(r^2 K^S_{rr}) - \frac{\partial}{\partial \mu}\left(\frac{ K^S_{r\theta}\sqrt{1-\mu^2}}{r}\right) + V_{\rm{sw}}+v_{{\rm dr},r}+v_{\rm HCS}\right]\Delta t +\nonumber\\
           &&    + \omega_r\sqrt{\frac{K^S_{rr}K^S_{\theta\theta}-(K^S_{r\theta})^2}{0.5 K^S_{\theta\theta}}\,\Delta t\,}   -  \omega_{\mu} K^S_{r\theta}\sqrt{\frac{2}{K^S_{\theta\theta}}\,\Delta t\,},
           \label{eq::app7}   \\
\Delta \mu &=& \left\{-\frac{1}{r^2} \frac{\partial}{\partial r}\left(r K^S_{\theta r}\sqrt{1-\mu^2}\right)+ \frac{\partial}{\partial \mu}\left[ \frac{K^S_{\theta \theta}(1-\mu^2)}{r^2}\right]-\frac{v_{{\rm dr},\theta}\sqrt{1-\mu^2}}{r}\right\}\Delta t +\nonumber\\
           &&  +\omega_{\mu}  \sqrt{\frac{2 K^S_{\theta \theta}(1-\mu^2)}{r^2}\, \Delta t\,}, \label{eq::app8}\\
\Delta T  &=&  - \frac{2}{3} \frac{\alpha_{\rm rel} V_{\rm sw} T}{r}\Delta t \label{eq::app9}.
\end{eqnarray}

%We refer to [\cite{jokipii77b}] and [\cite{Achterberg1992}] for a first introduction on how to use SDE approach on CR propagation.
%
\section*{Acknowledgements}
The authors express recognition for the contribution of Francesco Noventa to simulations regarding the heliosphere dimensions.~KK wishes to acknowledge VEGA grant agency project 2/0081/10 for support.~Finally, we acknowledge the use of NASA/GSFC's Space Physics Data Facility's OMNIWeb service, and OMNI data.
%
%\clearpage
%
%\vfill \eject

\end{document}